\documentclass[aps,prd,twocolumn,superscriptaddress,nofootinbib,floatfix,longbibliography]{revtex4-2}

\usepackage{amsmath,amssymb,bm}
\usepackage{graphicx}
\usepackage{ragged2e}
\usepackage{subfig}        
\usepackage{xcolor}
\usepackage[colorlinks=true,allcolors=blue]{hyperref}

\newcommand{\Cn}{C_n}

\newcommand{\kmsmpc}{\,\mathrm{km\,s^{-1}\,Mpc^{-1}}}
\usepackage{orcidlink}

\makeatletter
\long\def\@makecaption#1#2{%
  \vskip\abovecaptionskip
  \sbox\@tempboxa{#1: #2}%
  \ifdim \wd\@tempboxa > \linewidth
    \begin{center}
      \parbox{\linewidth}{\justifying #1: #2}
    \end{center}
  \else
    \global\@minipagefalse
    \hb@xt@\linewidth{\hskip 0pt #1: #2\hfil}%
  \fi
  \vskip\belowcaptionskip}
\makeatother

\begin{document}

\title{Early- and late-time constraints on Wald-Gauss-Bonnet topological
dark energy and implications for the $H_0$ and $S_8$ tensions}

\author{Stylianos A.\ Tsilioukas\,\orcidlink{0009-0003-3051-3405}}
\email{tsilioukas@sch.gr}
\affiliation{\mbox{National Observatory of Athens, Lofos Nymfon, 11852 Athens, 
Greece}}

\author{Fotios K. Anagnostopoulos}
\email{fotisanagn@uop.gr}
\affiliation{Department of Informatics and Telecommunications, University of Peloponnese, Karaiskaki 70, 22100, Tripoli, Greece}

\author{Spyros~Basilakos\,\orcidlink{0000-0001-5066-0259}}
\email{svasil@academyofathens.gr}
\affiliation{\mbox{National Observatory of Athens, Lofos Nymfon, 11852 Athens, 
Greece}}
\affiliation{Academy of Athens, Research Center for Astronomy and Applied 
Mathematics, Soranou Efesiou 4, 11527, Athens, Greece}
\affiliation{School of Sciences, European University Cyprus, Diogenes Street, 
Engomi, 1516 Nicosia, Cyprus}

\author{Andronikos Paliathanasis\,\orcidlink{0000-0002-9966-5517}}
\email{anpaliat@phys.uoa.gr}
\affiliation{Institute of Systems Science, Durban University of Technology, Durban 4000, South Africa}
\affiliation{Centre for Space Research, North-West University, Potchefstroom 2520, South Africa}
\affiliation{Centro de Investigaci\'on, Innovaci\'on y Creaci\'on (CIIC), Universidad Cat\'olica de Temuco, Temuco, Chile}
\affiliation{Departamento
de Ciencias Matem\'{a}ticas y F\'{\i}sicas,  Facultad de Ingenier\'{\i}a, Universidad Cat\'olica de Temuco, Temuco, Chile}
\affiliation{National Institute for Theoretical and Computational Sciences (NITheCS),
South Africa}

\author{Emmanuel N. Saridakis\,\orcidlink{0000-0003-1500-0874}}
\email{msaridak@noa.gr}
\affiliation{\mbox{National Observatory of Athens, Lofos Nymfon, 11852 Athens, 
Greece}}
\affiliation{CAS Key Laboratory for Researches in Galaxies and Cosmology, 
Department of Astronomy, \\
University of Science and Technology of China, Hefei, 
Anhui 230026, P.R. China}
\affiliation{\mbox{Departamento de Matem\'{a}ticas, Universidad Cat\'{o}lica 
del 
Norte, 
Avda.
Angamos 0610, Casilla 1280 Antofagasta, Chile}}

\begin{abstract}
The persistent $H_0$ and $S_8$ tensions motivate the search for new dark-energy
mechanisms capable of modifying the late-time expansion history while preserving
the successful early-Universe predictions of $\Lambda$CDM scenario. Wald-Gauss-Bonnet
(WGB) topological dark energy provides a physically motivated   realization of this possibility,
where the effective dark-energy sector emerges from cosmic horizon thermodynamics
and the black-hole formation and merger history. We present the first
early- and late-Universe analysis of WGB cosmology, implementing the model as an
effective fluid in a modified \texttt{CLASS} solver and constraining it against
CMB data from \textit{Planck}, ACT~DR6 and SPT-3G, DESI~DR2 BAO, and Pantheon+
supernovae.
While late-time data alone are consistent with $\Lambda$CDM, the full dataset
prefers a non-zero WGB contribution,
$\Cn=0.387^{+0.118}_{-0.072}$, corresponding to a $>3\sigma$ phantom-like
deviation and an improved fit ($\Delta\chi^2=-10.3$). The preferred solution raises
$H_0$ from $68.5$ to $69.8~\kmsmpc$, reducing the Hubble tension by
$\sim1\sigma$, at the cost of a mild increase in $S_8$. The reconstructed cosmological
observables show that WGB leaves the primary CMB almost unchanged while
enhancing lensing and small-scale clustering, revealing a characteristic
$H_0$-$S_8$ trade-off. WGB dark energy therefore emerges as a physically motivated and observationally viable late-time mechanism for partially alleviating the Hubble tension without introducing new early-Universe physics.
\end{abstract}

\maketitle

\section{Introduction}\label{sec:intro}

The $\Lambda$ Cold Dark Matter ($\Lambda$CDM) model has been remarkably
successful in describing the evolution and large-scale structure of the Universe.
Yet it continues to face challenges at both the theoretical and the observational
level. On the theoretical side, the non-renormalisability of general
relativity~\cite{AlvesBatista2023} and the cosmological-constant problem remain
unresolved. On the observational side, the growing precision of cosmological data
has revealed persistent tensions between the predictions of $\Lambda$CDM and
direct measurements of cosmological parameters~\cite{Abdalla2022}.

The most severe of these is the $H_0$ tension: the discrepancy between the
present-day Hubble constant inferred from early-Universe probes such as the
\textit{Planck} CMB combined with BAO, $H_0=67.27\pm0.60\kmsmpc$~\cite{Planck2018},
and its direct distance-ladder determination by SH0ES,
$H_0=73.04\pm1.04\kmsmpc$~\cite{Riess2022}, a discrepancy now exceeding the
$5\sigma$ level and resistant to explanation through local
systematics~\cite{Verde2019,DiValentino2021,RiessJWST2023,Cai:2026swf,Tiwari:2026pzk,Dainotti:2025qxz,RoyChoudhury:2026yyk}. A second challenge is
the $\sigma_8$ (or growth) tension, a difference in the amplitude of matter
clustering inferred from the CMB versus weak-lensing and large-scale-structure
surveys~\cite{Heymans2021,Zarrouk2018,Alam2017}. Notably, a series of recent
analyses indicate that this tension may have weakened: the complete
KiDS-Legacy cosmic-shear analysis shifts upward into $\lesssim\!1\sigma$
consistency with \textit{Planck}~\cite{WrightKiDSLegacy,StolznerKiDSLegacy}, a
recalibration of the DES~Year-3 astrophysical systematics raises $S_8$ by
${\sim}1.5\sigma$ toward the CMB value~\cite{Bigwood2025}, and a recent
compilation of the 2025 weak-lensing, clustering and growth measurements finds
them converging on a CMB-consistent amplitude, $S_8=0.819\pm0.007$
(${\sim}1.2\sigma$)~\cite{Wright:2025xka,Stolzner:2025htz}. Even so, the possibility of residual
inconsistencies across probes keeps open the question of whether new physics is
required. A broad range of extensions has been proposed to address these
tensions,  including early dark energy scenarios  
\cite{Poulin:2018cxd,Sakstein:2019fmf,Gogoi:2020qif,Niedermann:2020dwg,
Murgia:2020ryi,Seto:2021xua,Karwal:2021vpk}, modified 
recombination 
histories 
\cite{Liu:2019awo,Ye:2020btb,Lee:2022gzh,
Rashkovetskyi:2021rwg,Lynch:2024hzh,Mirpoorian:2024fka,
Jedamzik:2025cax,Pedrotti:2026dwj}, 
interacting dark sectors 
\cite{DiValentino:2017iww,An:2017crg,Yang:2018qmz,Yang:2018uae,Pan:2019jqh,
Pan:2019gop,Amirhashchi:2020qep,Gao:2021xnk,Yao:2022kub,Pan:2020bur,Paliathanasis:2026ymi,Zhang:2025dwu,Li:2026xaz,Li:2026asg}, entropic models  \cite{Basilakos:2023kvk,Yarahmadi:2024lzd,
Adhikary:2025khr,Yarahmadi:2025ujq,Li:2025vqt,Leizerovich:2026pfy,Luciano:2025ovj,Halder:2026wvg},
  modifications of gravity 
\cite{Khosravi:2017hfi,Nunes:2018xbm,El-Zant:2018bsc,Cai:2019bdh,Yan:2019gbw,
Escamilla-Rivera:2019ulu,Odintsov:2020qzd,Odintsov:2025egb,Odintsov:2026fr,Ballardini:2020iws,
Barker:2020gcp,Braglia:2020auw,Adi:2020qqf,
Petronikolou:2021shp,Adil:2021zxp,Paliathanasis:2025hjw,Nojiri:2022ski,Banerjee:2022ynv,Schiavone:2022wvq,Ren:2022aeo,Montani:2023xpd,
Boiza:2025xpn,Bouhmadi-Lopez:2026dte,Paliathanasis:2026vhi,DeSimone:2024lvy,Valletta:2025bgu}, etc (see 
\cite{CosmoVerseNetwork:2025alb} for a review).

In this work we investigate the observational viability of Wald-Gauss-Bonnet
(WGB) topological dark energy~\cite{Tsilioukas2024}, a physically motivated
dark-energy scenario arising from the application of the
spacetime-thermodynamics conjecture to the apparent horizon of the Universe.
Replacing the standard Bekenstein-Hawking entropy with the
Wald-Gauss-Bonnet entropy leads to modified Friedmann equations, assuming
that the apparent horizon is topologically linked to the black-hole horizons
contained within it. The resulting dark-energy sector is determined by the
cosmic history of black-hole formation and mergers, approximated through the
cosmic star-formation rate, thereby introducing an additional astrophysically
generated contribution to the cosmological constant. The model preserves the
standard thermal history of the Universe while naturally admitting both
quintessence- and phantom-like effective dark-energy behaviour. Since phantom
dynamics has been shown to alleviate the $H_0$ and $\sigma_8$
tensions~\cite{Abdalla2022,Heisenberg2023,Heisenberg2022}, WGB cosmology
provides a promising and physically motivated framework for addressing the
remaining cosmological tensions. Conceptually, it is related to, but physically
distinct from, the recently proposed topological dark-energy (TDE)
scenario~\cite{TsilioukasTDE2024,Tsilioukas:2024tjh,ANAGNOSTOPOULOS2026140626}: whereas TDE derives
dark energy from spacetime-foam topology, WGB links it to the formation and
merger of black-hole horizons.

The late-time phenomenology of WGB cosmology was established in the companion
analysis~\cite{Petronikolou:2025mlm}, where the model was confronted with
supernovae, cosmic chronometers and BAO observations and found to be
statistically compatible with $\Lambda$CDM. However, its confrontation with CMB
observations, which constitutes a decisive test of any cosmological model, was deliberately left
for future work. In this paper we perform the first comprehensive early- and
late-Universe analysis of WGB topological dark energy. We implement the WGB
background evolution as a dynamical dark-energy fluid in a modified version of
the \texttt{CLASS} Boltzmann solver and constrain the model, jointly with
$\Lambda$CDM one, using five combinations of CMB, BAO and supernova observations
within a Markov-chain Monte Carlo framework. Throughout this work we consider
the $\Lambda\neq0$ realisation (``Model~II''), which possesses an explicit
$\Lambda$CDM limit and a well-defined early-Universe behaviour. We therefore
assess whether a dark-energy sector generated by black-hole thermodynamics is
consistent with current cosmological observations and capable of alleviating
the remaining cosmological tensions. This analysis was carried out within the
CosmoVerse Compilation Group (CCG) initiative of the CosmoVerse
network~\cite{CosmoVerse2025}, and its results will be featured in the
forthcoming CosmoVerse white paper.

This paper is organised as follows. In Section \ref{sec:model} we review the WGB
cosmological framework and its implementation in the modified
\texttt{CLASS} Boltzmann solver. Section~\ref{sec:data} describes the datasets
and statistical methodology. Section~\ref{sec:results} presents the
cosmological constraints, model comparison and tension analysis. Finally, 
Section \ref{sec:recon} discusses the reconstructed observables, and
Section \ref{sec:conclusions} summarises our conclusions.

\section{Wald-Gauss-Bonnet cosmology}\label{sec:model}

In this section we briefly review the Wald-Gauss-Bonnet (WGB)
cosmological framework~\cite{Tsilioukas2024} and describe its
implementation in the \texttt{CLASS} Boltzmann solver. Augmenting the
Einstein-Hilbert action with the Gauss-Bonnet invariant
\begin{equation}
\mathcal{G}=R^2-4R_{\mu\nu}R^{\mu\nu}
+R_{\mu\nu\rho\sigma}R^{\mu\nu\rho\sigma},
\end{equation}
which  in four spacetime dimensions  is a topological invariant,   the action integral
\begin{equation}
S=\frac{1}{16\pi G}\int d^4x\,\sqrt{-g}\,
(R+\tilde\alpha\,\mathcal{G}),
\label{eq:action}
\end{equation}
provides the equations of motion of General Relatitivy, since in four
dimensions $\sqrt{-g}\,\mathcal{G}$ is a total derivative. The Gauss-Bonnet
term nevertheless leaves an imprint on the horizon entropy. For a
diffeomorphism-invariant Lagrangian $\mathcal{L}$, the Wald entropy is the
Noether charge associated with the horizon-generating Killing field, evaluated
on a spacelike cross-section $h$ of the horizon~\cite{Wald1993,IyerWald1994},
\begin{equation}
S_{\rm Wald}=-2\pi\oint_h
\frac{\partial\mathcal{L}}{\partial R_{\mu\nu\rho\sigma}}\,
\epsilon_{\mu\nu}\epsilon_{\rho\sigma}\,\sqrt{\sigma}\,d^2x ,
\label{eq:wald}
\end{equation}
where $\epsilon_{\mu\nu}$ is the binormal of $h$ (normalised as
$\epsilon_{\mu\nu}\epsilon^{\mu\nu}=-2$) and $\sigma$ the determinant of its
induced metric. Although $\mathcal{G}$ does not contribute to the bulk field
equations, its variation with respect to the Riemann tensor does not vanish,
and Eq.~(\ref{eq:wald}) restricted to the two-dimensional surface $h$ yields
\begin{equation}
S_{\rm WGB}=\frac{1}{4G}\oint_h\big(1+2\tilde\alpha\,\mathcal{R}\big)
\sqrt{\sigma}\,d^2x ,
\label{eq:swgb_int}
\end{equation}
with $\mathcal{R}$ the Ricci scalar of the induced metric on $h$. The second
term is the two-dimensional Euler density, so that by the Gauss-Bonnet theorem
$\oint_h\mathcal{R}\sqrt{\sigma}\,d^2x=4\pi\chi(h)$ and the Wald entropy
acquires a topological correction to the standard Bekenstein-Hawking area law,
\begin{equation}
S_{\rm WGB}
=\frac{A}{4G}
+\frac{2\pi\tilde\alpha}{G}\,\chi(h),
\label{eq:swgb}
\end{equation}
where $\tilde\alpha$ is the Gauss-Bonnet coupling,
$\chi(h)$ denotes the Euler characteristic of the horizon $h$, and
$A=4\pi r_h^2$ is the horizon area. The correction is therefore a pure
constant for a horizon of fixed topology, and it can affect the dynamics only
if $\chi(h)$ changes in time.

\subsection{Background evolution}\label{sec:background}

The derivation proceeds in two logically distinct steps. First, we apply the
gravity-thermodynamics conjecture with the entropy (\ref{eq:swgb}) to the
apparent horizon of a spatially flat FRW Universe, keeping $\chi$ an arbitrary
function of time; this step is exact within the
conjecture~\cite{Jacobson1995,CaiKim2005,AkbarCai2007,CaiCao2007}. Second, we
specify $\chi(t)$ through a phenomenological ansatz that links the horizon
topology to the black-hole population~\cite{Tsilioukas2024}.

\textit{Step (i): first law at the apparent horizon.} For a flat FRW metric
the apparent horizon has radius $\tilde r_A=1/H$, area
$A=4\pi\tilde r_A^2$ and temperature $T=1/(2\pi\tilde r_A)$. The energy of the
matter fluid crossing the horizon during $dt$ is~\cite{CaiKim2005}
\begin{equation}
-dE=A\,(\rho_m+p_m)\,H\tilde r_A\,dt=\frac{4\pi}{H^2}\,(\rho_m+p_m)\,dt ,
\label{eq:flux}
\end{equation}
while differentiating Eq.~(\ref{eq:swgb}) and using
$d\tilde r_A=-\dot H\,dt/H^2$ gives
\begin{align}
T\,dS_{\rm WGB}&=\frac{H}{2\pi}\left(\frac{2\pi\tilde r_A}{G}\,d\tilde r_A
+\frac{2\pi\tilde\alpha}{G}\,d\chi\right)\nonumber\\
&=\left(-\frac{\dot H}{G H^2}+\frac{\tilde\alpha H}{G}\,\dot\chi\right)dt .
\label{eq:TdS}
\end{align}
Imposing the first law, $-dE=T\,dS_{\rm WGB}$, we obtain the modified second
Friedmann equation,
\begin{equation}
\dot H=-4\pi G(\rho_m+p_m)+\tilde\alpha\,H^3\dot\chi .
\label{eq:firstlaw}
\end{equation}
Using the matter conservation equation, $\dot\rho_m=-3H(\rho_m+p_m)$, the
right-hand side of $\frac{d}{dt}H^2=2H\dot H$ becomes
$\frac{8\pi G}{3}\dot\rho_m+2\tilde\alpha H^4\dot\chi$, and a time integration
yields the modified first Friedmann equation,
\begin{equation}
H^2=\frac{8\pi G}{3}\rho_m+\frac{\Lambda}{3}
+2\tilde\alpha\int_0^t H^4\,\dot\chi\,dt' ,
\label{eq:integrated}
\end{equation}
in which the cosmological constant appears as an integration constant. For
$\tilde\alpha\to0$, or for a horizon of constant topology ($\dot\chi=0$),
Eq.~(\ref{eq:integrated}) reduces to the standard $\Lambda$CDM Friedmann
equation.

\textit{Step (ii): topology change from black-hole formation and mergers.}
Following Ref.~\cite{Tsilioukas2024}, we assume that the apparent horizon is
topologically linked to the black-hole horizons $h_i$ it encloses, so that
the Euler characteristic of the total boundary,
$\chi(\partial M)=\chi(H)+\sum_i\chi(h_i)$, is conserved~\cite{Liko2008,SarkarWall2011}.
Since every black-hole formation adds a spherical horizon
($\delta\chi(h)=+2$) and every merger removes one ($\delta\chi(h)=-2$), the
Euler characteristic of the apparent horizon changes as
\begin{equation}
\delta\chi=-2(\delta N_{\rm form}-\delta N_{\rm merg})\equiv-2\,\delta N .
\label{eq:dchi}
\end{equation}
This is the phenomenological input of the model: it is motivated by the
observation that the topological part of the Wald entropy is not monotonic
during black-hole mergers~\cite{Liko2008,SarkarWall2011}, and it renders
$\chi$ dynamical. The black-hole formation rate inside the apparent-horizon
volume $V_A=4\pi\tilde r_A^3/3$ is approximated through the Madau-Dickinson
cosmic star-formation rate~\cite{MadauDickinson2014},
\begin{equation}
\psi(z)=0.015\,
\frac{(1+z)^{2.7}}
{1+[(1+z)/2.9]^{5.6}}
\;\;M_\odot\,{\rm yr}^{-1}\,{\rm Mpc}^{-3},
\label{eq:sfr}
\end{equation}
so that $\dot N_{\rm form}=V_A f_{\rm BH}\,\psi/\langle m_{\rm prog}\rangle$
and $\dot N_{\rm merg}=f_{\rm bin}f_{\rm merge}\dot N_{\rm form}$, i.e.,
\begin{equation}
\dot N=C\,\frac{\psi(z)}{H^3},\quad
C\equiv
\frac{4\pi}{3}\,
\frac{(1-f_{\rm bin}f_{\rm merge})f_{\rm BH}}
{\langle m_{\rm prog}\rangle},
\label{eq:Cdef}
\end{equation}
where $f_{\rm BH}$ is the fraction of stellar mass in black-hole
progenitors with average mass $\langle m_{\rm prog}\rangle$, $f_{\rm bin}$ is
the fraction of the formed black holes that are in binary systems, and
$f_{\rm merge}$ is the fraction of them that eventually
merges~\cite{HegerWoosley2002,Fryer2012,Belczynski2016,Sana2012}.
Inserting Eqs.~(\ref{eq:dchi}) and (\ref{eq:Cdef}) into
Eq.~(\ref{eq:integrated}), the integrand becomes
$H^4\dot\chi=-2C\,H\psi$, and changing the integration variable with
$dt=-dz/[H(1+z)]$ (so that $t'\in[0,t]$ maps to $z'\in[z_i,z]$, with $z_i$ the
onset of star formation) gives
\begin{equation}
H^2(z)
=
H_0^2\Omega_{m0}(1+z)^3
+\frac{\Lambda}{3}
-
4\tilde\alpha C
\!\int_{z}^{z_i}
\frac{\psi(z')}{1+z'}
\,dz'.
\label{eq:friedmann}
\end{equation}

Since the astrophysical quantities enter only through the product
$\tilde\alpha C$, the model is effectively characterised by a single
parameter. To make it dimensionless, we introduce the reference
star-formation-rate density $\psi_\star\equiv1\,M_\odot\,{\rm yr}^{-1}\,{\rm Mpc}^{-3}$
and the normalised rate $\hat\psi(z)\equiv\psi(z)/\psi_\star$, i.e. the
numerical value of Eq.~(\ref{eq:sfr}), and define
\begin{equation}
\Cn\equiv
-\frac{\tilde\alpha C\,\psi_\star}{2H_0^2},
\label{eq:cn}
\end{equation}
which includes the combined phenomenological impact of the
substantial uncertainties associated with
$f_{\rm BH}$, $f_{\rm merge}$, $f_{\rm bin}$ and
$\langle m_{\rm prog}\rangle$, and in terms of which the topological term in
Eq.~(\ref{eq:friedmann}) reads
$+8\Cn H_0^2\int_z^{z_i}\hat\psi(z')/(1+z')\,dz'$. In all expressions below
written in terms of $\Cn$, the star-formation rate therefore enters through
the dimensionless $\hat\psi$, which is also the convention adopted in
\texttt{class\_wgb}.\footnote{Carrying out
explicitly the first-law step and the change of variable $t\to z$ fixes the
coefficient and sign of the topological term as in
Eq.~(\ref{eq:friedmann}); in Refs.~\cite{Tsilioukas2024,Petronikolou:2025mlm}
this term was quoted as $-8\tilde\alpha C\int_{z_i}^{z}\psi/(1+z')\,dz'$.
Because $\tilde\alpha$ and $C$ enter only through their product, the
difference is fully absorbed in the normalisation (\ref{eq:cn}), and all
expressions below, written in terms of $\Cn$, are unaffected. The only
consequence is interpretational: the phantom branch $\Cn>0$ corresponds to
$\tilde\alpha<0$, for which the topological entropy grows with the net
number of black holes, $\delta S_{\rm top}=-4\pi\tilde\alpha\,\delta N/G>0$.}
Unlike the companion late-time
analysis~\cite{Petronikolou:2025mlm}, which adopted a dimensionful
coupling, the normalization
(\ref{eq:cn}) renders $\Cn$ dimensionless and naturally of order
$\mathcal{O}(0.1)$, thereby decoupling it from the absolute scale of
$H_0$.

The framework admits two distinct realizations. In the $\Lambda=0$
case (``Model~I''), the WGB sector accounts for the entire dark-energy
budget. However, this realization possesses neither an explicit $\Lambda$CDM
limit nor a well-defined effective-fluid description at high redshift,
where $\rho_{\rm DE}\rightarrow0$ for
$z\gtrsim z_i$, rendering it unsuitable for an early-Universe
analysis. This case was investigated at late times in
Ref.~\cite{Petronikolou:2025mlm}.

In the present work we therefore focus on the $\Lambda\neq0$
realization (``Model~II''), for which the normalization condition
$H(z=0)=H_0$ applied to Eq.~(\ref{eq:friedmann}) fixes
\begin{equation}
\Lambda
=
3(1-\Omega_{m0})H_0^2
-
24\Cn H_0^2
\!\int_0^{z_i}
\frac{\hat\psi(z')}{1+z'}
\,dz',
\label{eq:lambda}
\end{equation}
where for $\tilde\alpha\rightarrow0$ the $\Lambda$CDM limit is recovered.
Collecting the $\Lambda$ and topological terms of Eq.~(\ref{eq:friedmann})
into an effective dark-energy density,
$H^2=H_0^2\Omega_{m0}(1+z)^3+\frac{8\pi G}{3}\rho_{\rm DE}$, and using
Eq.~(\ref{eq:lambda}) to combine
$\int_0^{z_i}-\int_z^{z_i}=\int_0^z$, we obtain
\begin{align}
\frac{8\pi G}{3}\rho_{\rm DE}(z)&=\frac{H_0^2}{3}\,D(z),\nonumber\\
D(z)&\equiv3(1-\Omega_{m0})-24\Cn\,I(z),
\label{eq:rhodez}
\end{align}
with $I(z)\equiv\int_0^z\hat\psi(z')/(1+z')\,dz'$. Since matter is separately
conserved, so is $\rho_{\rm DE}$, and its equation of state follows from
$w_{\rm DE}=-1+\frac{(1+z)}{3}\,d\ln\rho_{\rm DE}/dz$ together with
$dI/dz=\hat\psi/(1+z)$:
\begin{equation}
w_{\rm DE}(z)
=
-1
-
\frac{8\Cn\,\hat\psi(z)}
{3(1-\Omega_{m0})
-
24\Cn\displaystyle
\int_0^z
\frac{\hat\psi(z')}{1+z'}
\,dz'},
\label{eq:wde}
\end{equation}
so that positive (negative) values of $\Cn$ correspond to phantom
(quintessence) behaviour, while the limit $\Cn\rightarrow0$
smoothly recovers the $\Lambda$CDM model.

\subsection{Implementation in \texttt{CLASS}}\label{sec:class}

To evolve the model consistently through the early Universe, we implement the
WGB sector as a dynamical dark-energy fluid in
\texttt{CLASS}.\footnote{The modified code,
\texttt{class\_wgb}, is publicly available at
\url{https://github.com/stelios-tsilioukas/class_wgb}.}
Expressing the equation of state~(\ref{eq:wde}) in terms of the scale factor,
using $1+z=a^{-1}$ and $dz/(1+z)=-da/a$, yields the form required by the
background integrator, namely
\begin{equation}
w_{\rm DE}(a) = -1 - \frac{8\Cn\,\hat\psi(a)}
   {3(1-\Omega_{m0}) - 24\Cn\displaystyle\int_{a}^{1}\frac{\hat\psi(a')}{a'}\,da'}.
\label{eq:wdea}
\end{equation}

The dark-energy density follows from the background continuity equation,
$\rho_{\rm DE}'=-3\mathcal{H}(1+w_{\rm DE})\rho_{\rm DE}$, where
$\mathcal{H}=aH$ is the conformal Hubble parameter and a prime denotes a
derivative with respect to conformal time. Substituting
Eq.~(\ref{eq:wdea}) into the formal solution
$\rho_{\rm DE}(a)=\rho_{{\rm DE},0}
\exp\!\left[3\int_a^1(1+w_{\rm DE})\,da'/a'\right]$, and noting that the
numerator of $3(1+w_{\rm DE})/a'$ is precisely the derivative of the
denominator, $dD/da'=24\Cn\hat\psi(a')/a'$, the integral is a logarithm and
gives the closed form
\begin{align}
\frac{\rho_{\rm DE}(a)}{\rho_{{\rm DE},0}}&=\frac{D(a)}{D(1)},\nonumber\\
D(a)&=3(1-\Omega_{m0})-24\Cn\!\int_a^1\frac{\hat\psi(a')}{a'}\,da' ,
\label{eq:rhode}
\end{align}
in agreement with Eq.~(\ref{eq:rhodez}). In \texttt{CLASS} the integral
$\int_a^1\hat\psi/a'\,da'$ is tabulated once and splined, and $w_{\rm DE}$, its
derivative and $\rho_{\rm DE}$ are evaluated analytically from it.

The expansion rate is then
\begin{equation}
H(a)=H_0
\sqrt{\Omega_{r0}a^{-4}
+\Omega_{m0}a^{-3}
+\Omega_{{\rm DE},0}
\frac{\rho_{\rm DE}(a)}{\rho_{{\rm DE},0}}},
\label{eq:Ha}
\end{equation}
where the present-day dark-energy density is fixed by spatial flatness, i.e.
$\Omega_{{\rm DE},0}=1-\Omega_{m0}-\Omega_{r0}$. The ratio
$\rho_{\rm DE}(a)/\rho_{{\rm DE},0}$ therefore fully encodes the
backreaction of the WGB contribution on the expansion history and, through the
background evolution, on cosmological distance measures and the CMB anisotropy
spectra.

Equation~(\ref{eq:rhode}) makes the physical domain of the model explicit:
$\rho_{\rm DE}>0$ requires $D(a)>0$ at all times. For $\Cn<0$ this holds
identically, since $D$ grows toward the past. For $\Cn>0$, $D$ decreases
toward the past and saturates once star formation shuts off, at
$D_{\min}=3(1-\Omega_{m0})-24\Cn I_{\max}$ with
$I_{\max}=\int_0^1\hat\psi(a)/a\,da\simeq0.144$, so that the phantom branch is
bounded by
\begin{equation}
\Cn<\Cn^{\rm crit}=\frac{3(1-\Omega_{m0})}{24\,I_{\max}}\simeq0.87\,(1-\Omega_{m0})
\approx0.61 ,
\label{eq:cncrit}
\end{equation}
beyond which $\rho_{\rm DE}$ would cross zero at high redshift. We therefore
adopt the flat prior $\Cn\in[-10,\,0.55]$: the upper edge keeps the sampler
safely below the singular boundary~(\ref{eq:cncrit}) for all values of
$\Omega_{m0}$ allowed by the data, while the lower edge is wide enough not to
truncate any of the posteriors. Points with $D(a)\le0$ are in any case
rejected during the background integration.

A comment on the perturbation treatment is appropriate. Because the
Gauss-Bonnet invariant is purely topological in four dimensions, the action
(\ref{eq:action}) yields the same field equations as general relativity at both
the background and linear-perturbation levels, and introduces no additional
propagating degrees of freedom. The modified cosmological dynamics arises
instead from the gravity-thermodynamics conjecture together with the
phenomenological evolution of the horizon topology, which fix the background
expansion but do not by themselves determine the perturbations. The
effective-fluid description of the WGB sector at the perturbation level is
therefore an additional phenomenological assumption, which we now motivate.
At the background level, the modified Friedmann equations~(\ref{eq:firstlaw}) and
(\ref{eq:integrated}) are equivalent to the standard ones sourced by an
additional, separately conserved component $\rho_{\rm DE}$, and the matter
sector obeys the usual conservation and Euler equations. The perturbations of
this effective component were studied in detail in the companion
analysis~\cite{Petronikolou:2025mlm} within the effective-fluid
formalism of Ref.~\cite{Mehrabi:2015hva}, by solving the coupled system for
the matter and dark-energy density contrasts and the gravitational potential
for an arbitrary effective sound speed $c_e$. Two results of that analysis
justify the treatment adopted here. First, the effective Newton constant that
governs the growth of matter perturbations deviates from the standard one by
only $G_{\rm eff}/G_N-1\sim10^{-3}$, in contrast to genuine
modified-gravity theories such as $f(T)$ gravity, where the perturbation sector
generates a gravitational slip and scale-dependent $\mu$ and $\Sigma$
functions that can dominate cosmological constraints
\cite{Verma:2026ios}. Second, dark-energy clustering is negligible: varying
the sound speed between $c_e=0$ and $c_e=1$ changes $f\sigma_8$ by only
$\Delta f\sigma_8/f\sigma_8\sim10^{-6}$, more than an order of magnitude below
the present observational uncertainties. Both properties reflect the fact
that the dark-energy perturbations are sourced by the combination
$\Omega_{\rm DE}(1+w_{\rm DE})$, which remains small throughout the evolution:
at the best fit of our full dataset combination, $|1+w_{\rm DE}|$ reaches
${\sim}0.3$ only around $z\simeq2$, where $\Omega_{\rm DE}\lesssim0.06$, so
that $\Omega_{\rm DE}|1+w_{\rm DE}|\lesssim0.03$ at all redshifts.

Adopting a minimally coupled, smooth dark-energy fluid is thus motivated by
this previous perturbation analysis, which demonstrated negligible
dark-energy clustering and very small deviations in the effective Newton
constant. We implement its perturbations using the
generic parametrised post-Friedmann (PPF) scheme~\cite{HuSawicki2007,Fang:2008sn} of \texttt{CLASS}, which is
designed precisely for smooth dark-energy components with $w(a)\neq-1$: it
reproduces the behaviour of a minimally coupled fluid with relativistic sound
speed on sub-horizon scales while remaining regular as $w_{\rm DE}\to-1$,
where the standard fluid equations become ill-defined. This is the relevant
situation for WGB, whose equation of state approaches $-1$ both today and at
high redshift, where the star-formation source $\psi$ switches off. Since the
companion analysis has established that the predictions are insensitive to
the dark-energy sound speed, the specific closure adopted for the
dark-energy perturbations has no observable impact. Consequently, all
growth-sector predictions presented below, including CMB lensing, the matter
power spectrum, $\sigma_8$ and $f\sigma_8$, arise entirely from the modified
expansion history, while the perturbation dynamics is that of general
relativity with a smooth dark-energy component. In particular, $\sigma_8$ is computed by \texttt{CLASS}
from the linear matter power spectrum with the standard top-hat window of
radius $R=8\,h^{-1}{\rm Mpc}$, which is a model-independent convention; WGB
modifies $\sigma_8$ only through the background-driven growth of $\delta_m$
and through the value of $h$ entering $R$.

\section{Data and methodology}\label{sec:data}
 
We now proceed to  confront the WGB cosmology with current cosmological observations.
To isolate the impact of different cosmological probes, we consider
complementary combinations of early- and late-Universe datasets and
analyse them within a common Bayesian framework. This allows us to
constrain the model parameters, compare the statistical performance of
WGB scenario against $\Lambda$CDM paradigm, and quantify its impact on the $H_0$ and
$S_8$ tensions.

\subsection{Datasets}\label{sec:datasets}

We use five combinations of early- and late-time
probes, designed to isolate the role of cosmic-microwave-background information
and of the local distance-ladder calibration.

\textit{Cosmic microwave background (CMB-SPA).} Our CMB dataset combines low,
intermediate and high-multipole information from three experiments through the
differentiable-likelihood framework \texttt{candl}~\cite{Balkenhol:2024sbv}: the
\textit{Planck}~2018 low-$\ell$ temperature likelihood, $\ell\in[2,29]$
\cite{Planck2018}; the ACT~DR6 CMB-only TT/TE/EE
likelihood~\cite{AtacamaCosmologyTelescope:2025blo} together with the ACT~DR6
CMB-lensing reconstruction~\cite{ACT:2023kun}; and the SPT-3G data, comprising
the D1 TT/TE/EE (``TnE'') spectra and the two-year delensed-EE and lensing
($\phi\phi$) analysis~\cite{Camphuis:2025,Ge:2025}. We denote this combination CMB-SPA. The
ACT and SPT-3G lensing information is what endows the dataset with genuine
sensitivity to the late-time clustering amplitude, and therefore to $S_8$.

\textit{Baryon acoustic oscillations.} We use the DESI~DR2 BAO
measurements~\cite{DESI:2025} ($N_{\rm data}=13$ distance ratios across the BGS,
LRG, ELG, QSO and Ly$\alpha$ tracers).

\textit{Type Ia supernovae.} We employ the Pantheon+
compilation~\cite{ScolnicPantheonPlus} in two variants. ``PP'' uses the $1624$
supernovae \textit{without} the SH0ES calibration (the $77$ Cepheid-host
calibrators removed), constraining the expansion history through relative
distances alone. ``PPS'' uses the full Pantheon+\,\&\,SH0ES
sample~\cite{Brout:2022vxf} of $1701$ supernovae including the Cepheid-calibrated
hosts, thereby folding the local distance-ladder determination of $H_0$ into the
likelihood.

The five combinations and their data-point counts are summarised in
Table~\ref{tab:datasets}: (1)~PP+DESI and (4)~PPS+DESI are purely late-time and
isolate the geometric constraints; (2)~CMB-SPA is CMB-only; (3)~CMB-SPA+PP+DESI
adds early-time information without the distance-ladder prior; and
(5)~CMB-SPA+PPS+DESI, the full early+late combination including SH0ES, is our
headline dataset.

\begin{table}[ht!]
\caption{Dataset combinations and corresponding numbers of data points. The
CMB-SPA likelihood comprises $N_{\rm candl}=341$ \texttt{candl} bandpowers,
of which $145$ are from the ACT~DR6 TT/TE/EE and lensing likelihoods and
$196$ from the SPT-3G D1 TnE likelihood, together with $28$ \textit{Planck}
low-$\ell$ TT data points.
\label{tab:datasets}}
\begin{ruledtabular}\begin{tabular}{lc}
Combination & $N_{\rm data}$ \\ \hline
PP+DESI            & $1637$ \\
PPS+DESI           & $1714$ \\
CMB-SPA            & $369$ \\
CMB-SPA+PP+DESI    & $2006$ \\
CMB-SPA+PPS+DESI   & $2083$ \\
\end{tabular}\end{ruledtabular}
\end{table}

\subsection{MCMC and model comparison}\label{sec:mcmc}

The WGB background evolution and linear perturbations are computed with a
modified version of the \texttt{CLASS} Boltzmann solver~\cite{CLASS}, in which
the topological dark-energy sector is implemented as an effective fluid, as
described in subsection \ref{sec:class}. Nonlinear scales are modelled using
HMcode-2020~\cite{Mead:2020vgs}, and we assume a single massive neutrino with
$\sum m_\nu = 0.06$~eV. Posterior distributions are obtained with the
\texttt{Cobaya} sampler~\cite{Cobaya} (v3.6.2) using the Metropolis algorithm
with fast-parameter dragging, evolving four independent chains for each dataset
combination on a 32-core node. Moreover, convergence is assessed using the Gelman-Rubin
statistic, requiring $R-1<0.02$ for the parameter means and $R-1<0.2$ for the
95\% credible intervals; all ten runs satisfy both criteria. The chains are
subsequently analysed with \texttt{GetDist}~\cite{GetDist} after discarding the
first $30\%$ of each chain as burn-in. We sample the six baseline cosmological
parameters
$\{H_0,\,\omega_b,\,\omega_c,\,\ln(10^{10}A_s),\,n_s,\,\tau\}$,
together with the WGB parameter $\Cn$ and, for the CMB combinations, the
\texttt{candl} calibration and foreground nuisance parameters.

\begin{table}[ht]
\caption{Priors adopted in the Bayesian analysis. $\mathcal{U}(a,b)$ denotes a
uniform distribution on $[a,b]$, and $\mathcal{N}(\mu,\sigma)$ a Gaussian
distribution. The lower block lists the \texttt{candl} instrumental calibration
nuisance parameters, which are varied only for the CMB dataset combinations.
\label{tab:priors}}
\begin{ruledtabular}
\begin{tabular}{lc}
Parameter & Prior distribution \\ \hline
$C_n$               & $\mathcal{U}(-10,\,0.55)$ \\
$\Omega_b h^2$      & $\mathcal{U}(0.005,\,0.1)$ \\
$\Omega_c h^2$      & $\mathcal{U}(0.001,\,0.99)$ \\
$H_0$               & $\mathcal{U}(40,\,100)$ \\
$\log(10^{10}A_s)$  & $\mathcal{U}(1.61,\,3.91)$ \\
$n_s$               & $\mathcal{U}(0.8,\,1.2)$ \\
$\tau_{\rm reio}$   & Fixed to $0.06$ (late-time only); \\
                    & otherwise $\mathcal{N}(\mu=0.051,\,\sigma=0.006)$ \\ \hline
$A_{\rm planck}$    & $\mathcal{U}(0.5,\,1.5)$ \\
$P_{\rm ACT}$       & $\mathcal{U}(0.9,\,1.1)$ \\
$E_{\rm cal}$       & $\mathcal{U}(0.8,\,1.2)$ \\
$T_{\rm cal}$       & $\mathcal{N}(1.0,\,0.0036)$ \\
\end{tabular}
\end{ruledtabular}
\end{table}

\begin{table*}\caption{Marginalized parameter constraints (posterior means and
$68\%$ credible intervals) for the WGB and $\Lambda$CDM models.
Parameters marked with $^{\dagger}$ are prior dominated in the
late-time-only dataset combinations and therefore do not represent
meaningful constraints. For the CMB-only combination the $\Cn$ posterior
extends to the upper prior edge, and we quote the one-sided $68\%$ lower
limit ($^{\ddagger}$).
\label{tab:constraints}}
\centering
\subfloat[PP+DESI, CMB-SPA, and CMB-SPA+PP+DESI\label{tab:constraints_a}]{%
\begin{ruledtabular}\begin{tabular}{lcccccc}
 & \multicolumn{2}{c}{PP+DESI} & \multicolumn{2}{c}{CMB-SPA} & \multicolumn{2}{c}{CMB-SPA+PP+DESI} \\
Parameter & WGB & $\Lambda$CDM & WGB & $\Lambda$CDM & WGB & $\Lambda$CDM \\ \hline
$H_0$ & $68.64^{+9.29}_{-6.02}$ & $72.24^{+10.29}_{-4.12}$ & $68.18^{+2.34}_{-1.11}$ & $67.32\pm0.37$ & $68.92\pm0.51$ & $68.15\pm0.24$ \\
$\Omega_m$ & $0.284\pm0.017$ & $0.304\pm0.008$ & $0.309^{+0.011}_{-0.022}$ & $0.317\pm0.005$ & $0.300\pm0.004$ & $0.305\pm0.003$ \\
$\Omega_b h^2$ & $0.0264^{+0.0130}_{-0.0058}$ & $0.0268^{+0.0128}_{-0.0049}$ & $0.0224\pm0.0001$ & $0.0224\pm0.0001$ & $0.0224\pm0.0001$ & $0.0225\pm0.0001$ \\
$\Omega_c h^2$ & $0.1090^{+0.0209}_{-0.0296}$ & $0.1333^{+0.0310}_{-0.0195}$ & $0.1203\pm0.0010$ & $0.1204\pm0.0009$ & $0.1193\pm0.0008$ & $0.1184\pm0.0006$ \\
$n_s$ & $0.9997\pm0.1153^{\dagger}$ & $1.0023\pm0.1162^{\dagger}$ & $0.9692\pm0.0033$ & $0.9692\pm0.0032$ & $0.9716\pm0.0031$ & $0.9733\pm0.0028$ \\
$\ln(10^{10}A_s)$ & $2.762\pm0.661^{\dagger}$ & $2.736\pm0.655^{\dagger}$ & $3.047\pm0.011$ & $3.049\pm0.010$ & $3.052\pm0.010$ & $3.059\pm0.009$ \\
$\tau$ & $-$ & $-$ & $0.054\pm0.005$ & $0.055\pm0.005$ & $0.056\pm0.005$ & $0.059\pm0.005$ \\
$\sigma_8$ & ${0.658^{+0.153}_{-0.313}}^{\dagger}$ & ${0.782^{+0.172}_{-0.351}}^{\dagger}$ & $0.822^{+0.020}_{-0.010}$ & $0.815\pm0.004$ & $0.824\pm0.007$ & $0.813\pm0.004$ \\
$S_8$ & ${0.643^{+0.152}_{-0.311}}^{\dagger}$ & ${0.788^{+0.175}_{-0.353}}^{\dagger}$ & $0.834^{+0.010}_{-0.012}$ & $0.837\pm0.009$ & $0.824\pm0.007$ & $0.819\pm0.006$ \\
$r_{\rm drag}$ & $148.03^{+9.12}_{-21.40}$ & $141.53^{+6.32}_{-19.86}$ & $146.91\pm0.24$ & $146.90\pm0.23$ & $147.16\pm0.21$ & $147.35\pm0.17$ \\
$C_n$ & $-0.899\pm0.664$ & $-$ & $>0.04^{\ddagger}$ & $-$ & $0.212\pm0.118$ & $-$ \\
\end{tabular}\end{ruledtabular}}
\\[8pt]
\subfloat[PPS+DESI and CMB-SPA+PPS+DESI\label{tab:constraints_b}]{%
\begin{ruledtabular}\begin{tabular}{lccccc}
 & \multicolumn{2}{c}{PPS+DESI} & \multicolumn{2}{c}{CMB-SPA+PPS+DESI} \\
Parameter & WGB & $\Lambda$CDM & WGB & $\Lambda$CDM \\ \hline
$H_0$ & $73.60\pm0.99$ & $73.80\pm0.97$ & $69.82^{+0.45}_{-0.37}$ & $68.48\pm0.24$ \\
$\Omega_m$ & $0.283^{+0.013}_{-0.016}$ & $0.304\pm0.008$ & $0.293\pm0.004$ & $0.300\pm0.003$ \\
$\Omega_b h^2$ & $0.0337^{+0.0037}_{-0.0031}$ & $0.0285\pm0.0013$ & $0.0224\pm0.0001$ & $0.0225\pm0.0001$ \\
$\Omega_c h^2$ & $0.1191\pm0.0115$ & $0.1365\pm0.0053$ & $0.1196\pm0.0007$ & $0.1177\pm0.0006$ \\
$n_s$ & $1.0084\pm0.1180^{\dagger}$ & $1.0089\pm0.1158^{\dagger}$ & $0.9709\pm0.0029$ & $0.9748\pm0.0028$ \\
$\ln(10^{10}A_s)$ & $2.752\pm0.671^{\dagger}$ & $2.861\pm0.637^{\dagger}$ & $3.049\pm0.010$ & $3.063\pm0.009$ \\
$\tau$ & $-$ & $-$ & $0.055\pm0.006$ & $0.061\pm0.005$ \\
$\sigma_8$ & ${0.672^{+0.150}_{-0.317}}^{\dagger}$ & ${0.840^{+0.196}_{-0.367}}^{\dagger}$ & $0.834\pm0.007$ & $0.813\pm0.004$ \\
$S_8$ & ${0.654^{+0.149}_{-0.311}}^{\dagger}$ & ${0.846^{+0.196}_{-0.360}}^{\dagger}$ & $0.824\pm0.007$ & $0.813\pm0.006$ \\
$r_{\rm drag}$ & $136.30\pm1.98$ & $136.91\pm1.99$ & $147.06\pm0.20$ & $147.49\pm0.17$ \\
$C_n$ & $-0.957^{+0.517}_{-0.621}$ & $-$ & $0.387^{+0.118}_{-0.072}$ & $-$ \\
\end{tabular}\end{ruledtabular}}
\end{table*}

We adopt flat priors throughout, including
$\Cn\in[-10,0.55]$, as justified in subsection \ref{sec:class}, and the full set of priors is
summarised in Table~\ref{tab:priors}. Since our CMB compilation includes only
the \textit{Planck} low-$\ell$ temperature likelihood and no large-scale
polarisation, the optical depth is not directly constrained by the data.
Following the baseline SPT-3G~D1 analysis~\cite{Camphuis:2025}, we therefore
impose a Gaussian prior,
$\tau=0.051\pm0.006$, for all CMB combinations.

For model comparison we evaluate the best-fit statistic,
$\chi^2_{\rm min}\equiv-2\ln\mathcal{L}_{\rm max}$, at the
maximum-a-posteriori (MAP) point, located with a BOBYQA minimisation seeded
from the converged chains. For the combinations that drive our conclusions
(Runs~3 and~5, and the SH0ES-augmented run used for $Q_{\rm DMAP}$ below) we
verified the minima with $16$ additional BOBYQA searches from random starting
points and a tighter convergence threshold. The lowest minima found agree
with the chain-seeded ones to within $0.2$ in $-\ln\mathcal{L}$ (to within
$0.02$ for both models in Run~5); we adopt the lowest value found in each
case, so that the quoted $\Delta\chi^2$ are accurate to a few tenths. We then compute the Akaike (AIC), Bayesian (BIC) and
Deviance (DevIC) information criteria
\cite{Liddle2007,KassRaftery1995}. Following standard practice, we adopt the
small-sample corrected AIC,
\begin{align}
\mathrm{AIC} &= -2\ln\mathcal{L}_{\rm max} + 2k
   + \frac{2k(k+1)}{N_{\rm tot}-k-1},\\
\mathrm{BIC} &= -2\ln\mathcal{L}_{\rm max} + k\ln N_{\rm tot},\\
\mathrm{DevIC} &= D(\overline{\phi}) + 2C_B,\qquad
   C_B = \overline{D(\phi)} - D(\overline{\phi}),
\end{align}
where $D(\phi)=-2\ln\mathcal{L}(\phi)$ is the deviance and
$\overline{\phi}$ denotes the posterior mean. Here
$k=\nu+\nu_i$ is the total number of sampled parameters
($\nu$ cosmological/model parameters and $\nu_i$ nuisance parameters), while
$N_{\rm tot}$ is the total number of data points. For the dataset combinations
considered here the correction term is negligible, so that
$\mathrm{AIC}\simeq-2\ln\mathcal{L}_{\rm max}+2k$.

Models are ranked using the relative differences
$\Delta\mathrm{IC}=\mathrm{IC}_{\rm model}-\mathrm{IC}_{\rm min}$, interpreted
according to the Jeffreys scale~\cite{KassRaftery1995}: values
$\Delta\mathrm{IC}\le2$ indicate statistical compatibility with the preferred
model, $2<\Delta\mathrm{IC}<6$ moderate evidence against it,
$6\le\Delta\mathrm{IC}<10$ strong evidence, and
$\Delta\mathrm{IC}\ge10$ decisive evidence. Since WGB differs from
$\Lambda$CDM by a single additional parameter, $\Cn$ ($\Delta k=1$), one has
$\Delta\mathrm{AIC}\simeq\Delta\chi^2+2$.

Throughout, $\chi^2\equiv-2\ln\mathcal{L}$ denotes the likelihood-only
(data) contribution, $\chi^2_{\rm tot}=\sum_p\chi^2_p$, summed over the
individual likelihoods, which are mutually independent; the \texttt{candl}
covariance between ACT, SPT-3G and \textit{Planck} is accounted for within
each \texttt{candl} likelihood block. The prior terms of the log-posterior
are excluded from all information criteria: for a flat prior they contribute
only the normalisation $2\ln\Delta\theta$ of each prior range, an artefact of
the chosen prior width rather than a property of the fit.

The deviance at the posterior mean, $D(\overline{\phi})$, is obtained by
re-evaluating the full likelihood at the posterior-mean parameters, while
$\overline{D(\phi)}$ is the chain average of $\chi^2$. Because the absolute
deviance contains an arbitrary normalisation constant that cancels in model
comparisons, we report only the difference $\Delta\mathrm{DevIC}$.

Finally, we quantify the $H_0$ tension between our CMB+BAO+SNIa constraints and
the local distance ladder using the difference-of-maximum-a-posteriori
estimator, $Q_{\rm DMAP}$~\cite{Raveri:2018wln}, which has become the standard
metric for comparing proposed resolutions of the Hubble tension. We adopt the
uncalibrated dataset combination
$D={}$CMB-SPA+PP+DESI and incorporate the SH0ES measurement,
$H_0=73.04\pm1.04\kmsmpc$~\cite{Riess2022}, as a Gaussian likelihood on the
Hubble constant. Since the \texttt{sn.pantheonplus} likelihood analytically
marginalises over the supernova absolute magnitude, the SH0ES constraint can be
added directly to the likelihood. We then evaluate
\begin{equation}
Q_{\rm DMAP}
=
\sqrt{\chi^2_{\rm MAP}(D+{\rm SH0ES})
-\chi^2_{\rm MAP}(D)},
\end{equation}
using BOBYQA minimisations with and without the SH0ES contribution. Because
only a single parameter is shared between the two datasets, $Q_{\rm DMAP}$ may
be interpreted directly as an effective tension significance. Throughout, we
use the data $\chi^2$ (the sum of the individual likelihood contributions), so
that the $\tau$ and nuisance-parameter priors cancel in the difference. As
cross-checks, we also compute the Gaussian estimator,
$T=|x_1-x_2|/\sqrt{\sigma_1^2+\sigma_2^2}$, while the Gaussian estimator applied to $S_8$
provides a complementary measure of the residual growth tension. The CMB-only
combination (Run~2) is excluded from all tension metrics, since its
prior-limited $\Cn$ posterior renders such estimates physically
uninformative.

\section{Results}\label{sec:results}

We now test the WGB scenario with cosmological observations to assess
whether it provides a statistically significant improvement over
$\Lambda$CDM while alleviating the $H_0$ and $S_8$ tensions. Throughout this
section, Runs~1-5 refer to the five dataset combinations listed in
Table~\ref{tab:datasets}.

\begin{figure}[ht!]
  \centering
  \includegraphics[width=\columnwidth]{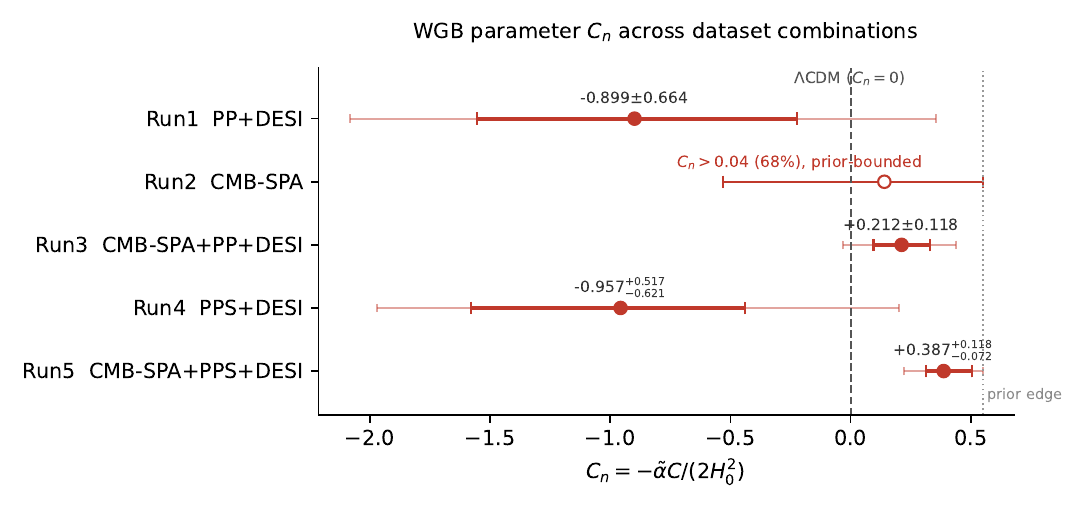}\caption{{\it
Marginalised $68\%$ and $95\%$ credible intervals for the WGB amplitude
$\Cn$ [Eq.~(\ref{eq:cn})] obtained from the five dataset combinations.
The late-time-only analyses (Runs~1 and~4) lean toward the quintessence side
but remain consistent with $\Cn=0$ at $95\%$ credibility, while the CMB-only
analysis (Run~2) is dominated by the $\Cn$-$H_0$ degeneracy and extends to the
upper edge of the prior. In contrast, combining CMB and late-time observations
favours a positive (phantom-side) WGB contribution, with the full dataset
combination (Run~5) yielding a more than $3\sigma$ preference for $\Cn>0$.
\label{fig:cnforest}}}
\end{figure}

\subsection{Parameter constraints}\label{sec:constraints}

The marginalized parameter constraints are summarised in
Table~\ref{tab:constraints}, while the corresponding measurements of the WGB
amplitude $\Cn$ are illustrated in Fig.~\ref{fig:cnforest}. The central result
is the constraint on $\Cn$. For the full early+late dataset combination
(Run~5, CMB-SPA+PPS+DESI), we obtain
$\Cn = 0.387^{+0.118}_{-0.072}$, corresponding to a more than
$3\sigma$ preference for a non-trivial phantom-side WGB contribution: the
$95\%$ lower limit is $\Cn>0.22$ and no posterior sample reaches $\Cn\le0$.
The asymmetric error reflects the upper prior edge, $\Cn=0.55$, which is
reached by the upper tail of the posterior ($12\%$ of the samples lie above
$\Cn=0.5$). Since $\Cn$ cannot exceed the physical bound
(\ref{eq:cncrit}), this truncation is largely physical; as a measure of its
impact, removing the tail $\Cn>0.5$ altogether shifts the derived $H_0$ and
$S_8$ by only $0.15\sigma$ and $0.08\sigma$, respectively. The CMB+late combination without the SH0ES
calibration (Run~3) yields the weaker constraint $\Cn = 0.212\pm0.118$,
corresponding to a ${\sim}2\sigma$ preference ($P(\Cn<0)=4\%$). The
late-time-only combinations (Runs~1 and~4) instead lean toward the
quintessence branch, $\Cn=-0.90\pm0.66$ and $-0.96^{+0.52}_{-0.62}$, but remain
consistent with $\Cn=0$ at $95\%$ credibility: without the CMB they constrain
$\Cn$ only weakly, through the shape of the low-redshift distance-redshift
relation. The CMB-only analysis (Run~2) represents a special case: the nearly
perfect $\Cn$-$H_0$ degeneracy ($\rho\simeq0.97$) drives $\Cn$ toward the upper
prior boundary, so the apparent displacement reflects the degeneracy direction
rather than evidence for a non-zero WGB contribution.

\begin{table*}[ht]\caption{Information-criterion comparison between the WGB and
$\Lambda$CDM models, defined as
$\Delta\equiv\mathrm{IC}_{\rm WGB}-\mathrm{IC}_{\Lambda\rm CDM}$, with
negative values favouring WGB. All criteria use the likelihood-only
$\chi^2$ (see subsection \ref{sec:mcmc}). The BIC is computed using the total
number of data points listed in Table~\ref{tab:datasets}. Only the full
dataset combination (CMB-SPA+PPS+DESI) shows a preference for WGB.
\label{tab:ic}}
\begin{ruledtabular}\begin{tabular}{lcccc}
Run & $\Delta\chi^2$ & $\Delta$AIC & $\Delta$BIC & $\Delta$DevIC \\ \hline
PP+DESI & $-2.29$ & $-0.29$ & $+5.11$ & $-0.67$ \\
CMB-SPA & $-1.07$ & $+0.93$ & $+4.84$ & $-1.72$ \\
CMB-SPA+PP+DESI & $-1.79$ & $+0.21$ & $+5.81$ & $-0.03$ \\
PPS+DESI & $-2.29$ & $-0.29$ & $+5.16$ & $-0.90$ \\
CMB-SPA+PPS+DESI & $-10.25$ & $-8.25$ & $-2.61$ & $-9.28$ \\
\end{tabular}\end{ruledtabular}\end{table*}

\begin{table*}[ht]
\caption{Best-fit $\chi^2_{\rm min}$ values at the MAP point, decomposed by
likelihood, for the two fully constrained dataset combinations; the last row
gives their sum, $\chi^2_{\rm tot}$. For Run~5, negative WGB$-\Lambda$CDM
differences identify the likelihoods favouring WGB (Pantheon+\,\&\,SH0ES, the
SPT-3G blocks and the ACT CMB-only spectra), whereas positive differences
identify the datasets that prefer $\Lambda$CDM (chiefly DESI~DR2 BAO).
\label{tab:chi2}}
\begin{ruledtabular}\begin{tabular}{lcccc}
 & \multicolumn{2}{c}{CMB-SPA+PP+DESI} & \multicolumn{2}{c}{CMB-SPA+PPS+DESI} \\
Likelihood & WGB & $\Lambda$CDM & WGB & $\Lambda$CDM \\ \hline
Pantheon+ (\&SH0ES) & $-$ & $-$ & 1479.44 & 1484.70 \\
Pantheon+ & 1408.13 & 1405.16 & $-$ & $-$ \\
DESI DR2 BAO & 13.80 & 15.02 & 13.79 & 11.65 \\
Planck low-$\ell$ TT & 22.26 & 22.07 & 22.18 & 21.83 \\
ACT DR6 (PlanckActCut) & 220.42 & 220.29 & 220.87 & 221.14 \\
ACT DR6 (CMBonly) & 159.02 & 159.91 & 158.66 & 160.26 \\
ACT DR6 lensing & 20.16 & 19.86 & 20.05 & 19.91 \\
SPT-3G D1 TnE & 160.74 & 162.31 & 160.74 & 163.64 \\
SPT-3G MUSE $\phi\phi$+EE & 21.56 & 23.25 & 21.49 & 24.36 \\
\hline
$\chi^2_{\rm tot}$ & 2026.08 & 2027.87 & 2097.23 & 2107.48 \\
\end{tabular}\end{ruledtabular}
\end{table*}

The preference for positive values of $\Cn$ in the CMB combinations propagates
directly to the late-time cosmological parameters. For the full dataset
combination (Run~5), WGB increases the Hubble constant from the
$\Lambda$CDM value of $H_0=68.48\kmsmpc$ to $69.82\kmsmpc$, accompanied by a
lower matter density ($\Omega_m=0.293$ instead of $0.300$) and a slightly
higher clustering amplitude
($\sigma_8=0.834$, $S_8=0.824$, compared with $0.813$ in $\Lambda$CDM).
As expected for the phantom branch, the same mechanism that raises $H_0$ also
enhances the growth of structure. We stress that, for the two purely
late-time runs, the primordial amplitude lies along a nearly flat likelihood
direction. Consequently, the parameters
$\{\ln(10^{10}A_s),\,\sigma_8,\,S_8,\,\tau,\,n_s\}$ are prior dominated and
are therefore flagged as such in Table~\ref{tab:constraints} rather than
quoted as meaningful constraints. We also note that, unlike $f(T)$ Model~II,
where the combined analysis drives the optical depth toward the edge of its
prior and into an unphysical region~\cite{Verma:2026ios}, the WGB posterior
remains comfortably within the imposed prior,
$\tau=0.055\pm0.006$, indicating no tension with the assumed
reionisation history.

As an independent consistency check, the physical baryon density inferred from
the calibrated late-time combination alone (PPS+DESI) can be compared with the
Big-Bang nucleosynthesis (BBN) determination,
$\Omega_b h^2\simeq0.0222$-$0.0223$~\cite{Cooke:2018}. We obtain
$\Omega_b h^2=0.0337^{+0.0037}_{-0.0031}$ for WGB and
$0.0285\pm0.0013$ for $\Lambda$CDM, implying tensions of approximately
$3.7\sigma$ and $4.8\sigma$, respectively. This reflects the well-known
preference of uncalibrated late-time distance measurements for a reduced sound
horizon once the SH0ES calibration fixes the absolute distance scale. Unlike
$f(T)$ gravity, where this test strongly discriminates between the two
Lambert-$W$ branches~\cite{Verma:2026ios}, this test does not discriminate
between WGB and $\Lambda$CDM, both being in comparable tension with BBN. As expected, consistency
with the BBN determination is restored for all dataset combinations that
include CMB observations.

\subsection{Model comparison}
\label{sec:modelcomp}

The model-comparison statistics are summarised in
Table~\ref{tab:ic}. They show a consistent picture: the additional WGB
parameter is favoured only when both early- and late-Universe observations,
including the SH0ES calibration, are analysed jointly. For the full dataset
combination (Run~5), the fit improves by
$\Delta\chi^2=-10.25$, yielding $\Delta\mathrm{AIC}=-8.25$ and
$\Delta\mathrm{DevIC}=-9.28$, corresponding to a strong preference for
WGB according to the Jeffreys scale. The effective number of constrained
parameters, $p_D=\overline{D(\phi)}-D(\overline{\phi})$, is $9.2$ for WGB and
$8.5$ for $\Lambda$CDM in this combination, i.e. the extra parameter is
genuinely constrained by the data, whereas in the remaining combinations the
DevIC differences are all below $2$ in magnitude. This is the only dataset combination for
which such a preference emerges. For all other runs the data remain compatible
with $\Lambda$CDM: the late-time, CMB-only and CMB+PP combinations give
$|\Delta\mathrm{AIC}|<1$, while the Bayesian information criterion, with its
stronger large-sample penalty, disfavors the additional parameter whenever it
is not well constrained ($\Delta\mathrm{BIC}=+4.8$ to $+5.8$ for Runs~1-4).
Remarkably, even the BIC favours WGB for the full combination,
$\Delta\mathrm{BIC}=-2.61$, the improvement in fit exceeding the
single-parameter penalty $\ln N_{\rm tot}=7.6$ over $N_{\rm tot}=2083$ data
points; since both minima are determined to within a few hundredths
(subsection~\ref{sec:mcmc}), this conclusion is robust. In absolute
terms, both models provide good fits, with reduced
$\chi^2\simeq0.87$ for the late-time combinations and
$\chi^2\simeq1$ whenever CMB data are included.

\begin{figure}[t]
  \centering
  \includegraphics[width=\columnwidth]{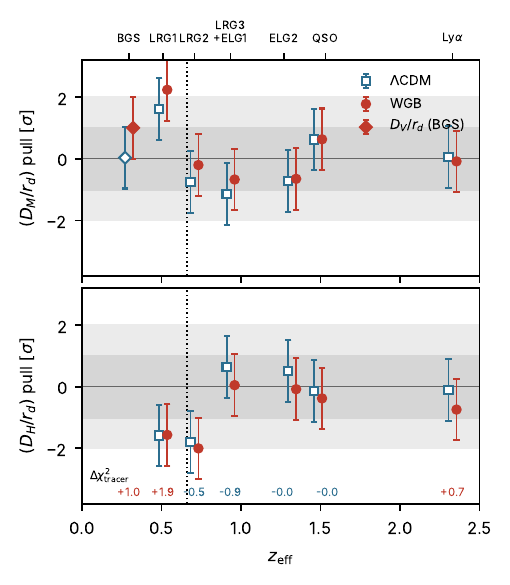}
  \caption{{\it
DESI DR2 BAO residuals, $({\rm data}-{\rm model})/\sigma$, evaluated at the
MAP of the full dataset combination (Run~5) for $\Lambda$CDM (open squares)
and WGB (filled circles). The upper panel shows $D_M/r_d$, with the BGS
measurement of $D_V/r_d$ indicated by the diamond, while the lower panel
shows $D_H/r_d$. The shaded bands denote the $\pm1\sigma$ and $\pm2\sigma$
regions, and the dotted vertical line marks the redshift,
$z\simeq0.66$, at which $H_{\rm WGB}$ crosses
$H_{\Lambda{\rm CDM}}$ (Fig.~\ref{fig:background}b). The numbers below the
panels give the per-tracer $\Delta\chi^2$ (WGB$-\Lambda$CDM), computed using
the full $D_M$-$D_H$ covariance matrix. The additional BAO cost of the WGB
model is concentrated at $z<0.66$ (BGS and LRG1, $\Delta\chi^2=+2.9$ in
total), where the enhanced late-time expansion rate predicts shorter distances
than preferred by these measurements.
\label{fig:desi}}}
\end{figure}

The per-likelihood decomposition, presented in
Table~\ref{tab:chi2}, identifies the origin of the Run~5 preference, while the
corresponding DESI BAO residuals are shown in
Fig.~\ref{fig:desi}. The improvement is driven primarily by the
Pantheon+\,\&\,SH0ES supernova likelihood
($\Delta\chi^2=-5.26$), together with the SPT-3G TnE likelihood
($-2.90$), the SPT-3G MUSE $\phi\phi$+EE ($-2.87$) and the ACT~DR6
temperature and polarization spectra ($-1.60$). These gains are partly offset
by the DESI~DR2 BAO data ($+2.14$), whose residuals reveal the expected modest
deterioration associated with the higher-$H_0$ solution, while the remaining
CMB blocks, including ACT~DR6 lensing ($+0.14$), are essentially neutral. In
the uncalibrated combination (Run~3) the same CMB blocks favour WGB
($-4.15$ in total from the SPT-3G and ACT spectra), whereas the Pantheon+
supernovae without the SH0ES calibration now mildly prefer $\Lambda$CDM
($+2.97$), so that the net improvement is only $\Delta\chi^2=-1.79$.

Since $\Lambda$CDM is nested within WGB at $\Cn=0$ and the prior on $\Cn$ is
separable and uniform, the Bayes factor can in principle be obtained directly
from the Savage-Dickey density ratio,
$B_{01}=p(\Cn=0\,|\,d)/\pi(\Cn=0)$, with $\pi(\Cn=0)=1/10.55$ for the adopted
flat prior. This estimator is reliable only where the marginal posterior is
well sampled at $\Cn=0$. For Runs~1, 3 and 4 it gives
$\ln B_{10}\simeq-1.1$, $-2.0$ and $-0.6$, respectively, stable to
$\pm0.2$ across kernel-density, histogram and skew-normal estimates of
$p(\Cn=0\,|\,d)$. For the full combination (Run~5), on the other hand, no
posterior sample reaches $\Cn\le0$, and the density at $\Cn=0$ can only be
extrapolated; the different estimators then range between $\ln B_{10}\simeq1.8$
and $3.8$, all favouring WGB. We therefore rely on the direct evidence
computation of subsection~\ref{sec:evidence} for Run~5, using the
Savage-Dickey values of Runs~1, 3 and 4 as an independent validation of that
method. Since the Bayes factor scales inversely with the prior width, the
deliberately broad flat prior adopted here renders these results conservative.
Any tighter physically motivated prior on
$\Cn$, for example derived from improved constraints on the astrophysical
factor $C$ [Eq.~(\ref{eq:Cdef})], would strengthen the evidence further.

\begin{figure}[ht]
  \centering
  \includegraphics[width=\columnwidth]{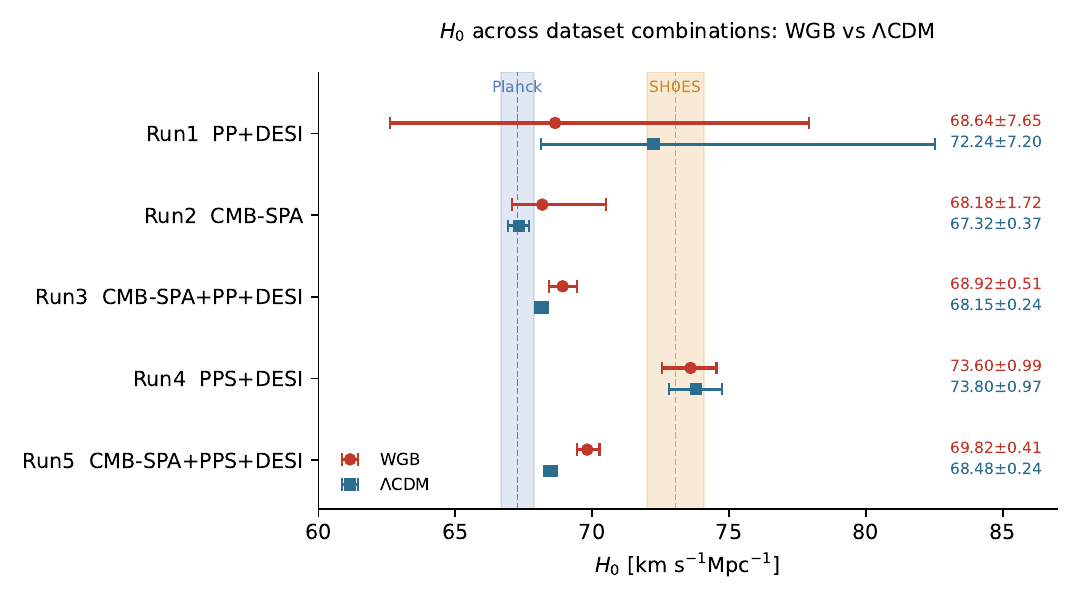}
  \caption{{\it
Marginalized $68\%$ and $95\%$ credible intervals for the Hubble constant
$H_0$ inferred from the five dataset combinations for the WGB and
$\Lambda$CDM models. The horizontal shaded bands indicate the
\textit{Planck} and SH0ES~\cite{Riess2022} determinations for reference.
Relative to $\Lambda$CDM, the WGB model systematically shifts the CMB-based
constraints toward higher values of $H_0$, thereby partially alleviating the
distance-ladder tension.
\label{fig:h0forest}}}
\end{figure}

\subsection{Bayesian evidence: an independent cross-check}\label{sec:evidence}

The Savage-Dickey ratio above is a convenient shortcut, but it relies on the
nesting of $\Lambda$CDM at $\Cn=0$, on the separability of the prior and on a
one-dimensional marginal density whose value at $\Cn=0$ cannot be estimated
reliably when that point lies far in the tail, as in Run~5. As a fully
independent method, not restricted to these assumptions, we compute the
Bayesian evidence $Z=\int\mathcal{L}(\theta)\,\pi(\theta)\,d\theta$ directly
from the existing chains using the $k$-th-nearest-neighbour estimator of
Ref.~\cite{Heavens:2017afc}, implemented in the public
\texttt{MCEvidence} code.\footnote{\url{https://github.com/yabebalFantaye/MCEvidence}}
The method estimates the local posterior density at each chain sample from the
distance to its $k$th nearest neighbour in parameter space, requiring no new
sampling and no assumption of nested models. Because our optical-depth and
\texttt{candl} calibration priors are Gaussian rather than flat, we supply the
unnormalised log-posterior (\texttt{cobaya}'s native $-\ln[\mathcal{L}\pi]$
column) directly as the input and fix the nominal prior-volume parameter to
unity, which recovers the estimator in its fully general form for arbitrary
prior shape. We validated this procedure against an analytic multivariate
Gaussian likelihood-prior pair, recovering the known evidence to within
$0.002$ in $\ln Z$. Table~\ref{tab:evidence} reports $\ln Z$ for both models
and the resulting Bayes factor, $\ln B_{10}=\ln Z_{\rm WGB}-\ln Z_{\Lambda{\rm
CDM}}$, in each of the five combinations. We quote the $k=4$ estimates; across the
$k$-NN scan ($k=1$-$4$) the Bayes factors vary by less than $0.4$, the
Run~5 value decreasing monotonically from $3.6$ to $3.2$, so that the quoted
value is the conservative one.

\begin{table}[t]
\caption{Bayesian evidence and Bayes factors from the $k$-NN estimator of
Ref.~\cite{Heavens:2017afc}, computed directly on the existing chains (no new
sampling). $\ln B_{10}=\ln Z_{\rm WGB}-\ln Z_{\Lambda{\rm CDM}}$; positive
favours WGB. $\dim$ lists the number of sampled parameters
(WGB/$\Lambda$CDM). $^{\dagger}$In Run~2 the $\Cn$ posterior extends to the
upper prior edge (see subsection \ref{sec:constraints}).\label{tab:evidence}}
\begin{ruledtabular}
\begin{tabular}{lcccc}
Run & $\dim$ & $\ln Z_{\rm WGB}$ & $\ln Z_{\Lambda{\rm CDM}}$ & $\ln B_{10}$ \\ \hline
PP+DESI                & 6/5   & $-718.09$  & $-717.23$  & $-0.87$ \\
CMB-SPA$^{\dagger}$     & 11/10 & $-343.77$  & $-343.65$  & $-0.11$ \\
CMB-SPA+PP+DESI         & 11/10 & $-1058.48$ & $-1056.70$ & $-1.78$ \\
PPS+DESI                & 6/5   & $-744.92$  & $-744.18$  & $-0.74$ \\
CMB-SPA+PPS+DESI        & 11/10 & $-1093.84$ & $-1097.05$ & $+3.22$ \\
\end{tabular}
\end{ruledtabular}
\end{table}

The picture is coherent across every combination and corroborates the rest of
the model-comparison analysis. In the four combinations where $\Cn$ is
statistically consistent with zero (Runs~1-4) the evidence is neutral to
mildly favouring $\Lambda$CDM, $\ln B_{10}=-0.1$ to $-1.8$, the expected Occam
penalty for a weakly constrained extra parameter. For Runs~1, 3 and 4 these
values agree to within $0.25$ with the Savage-Dickey estimates of
subsection~\ref{sec:modelcomp}, despite the two methods sharing no common
machinery, one a one-dimensional density at $\Cn=0$, the other a full
$k$-NN volume estimate in up to $11$ dimensions. In the full combination
(Run~5) the evidence independently favours WGB, $\ln B_{10}=3.22$,
corresponding to posterior odds of ${\sim}25{:}1$, i.e. strong evidence on the
Jeffreys scale. Expressed as $2\ln B_{10}=+6.4$, on the same numerical scale
as $\Delta$AIC by construction, the evidence lies between
$\Delta\mathrm{AIC}=-8.3$ and $\Delta\mathrm{BIC}=-2.6$, so that all three
criteria consistently favour WGB for the full dataset combination.

\subsection{The $H_0$ and $S_8$ tensions}\label{sec:tensions}

The impact of WGB on the inferred Hubble constant is illustrated in
Fig.~\ref{fig:h0forest}. The upward shift in $H_0$ partially alleviates the
tension with the SH0ES distance-ladder measurement,
$H_0=73.04\pm1.04\kmsmpc$~\cite{Riess2022}.

\begin{figure}[ht]
  \centering
  \includegraphics[width=\columnwidth]{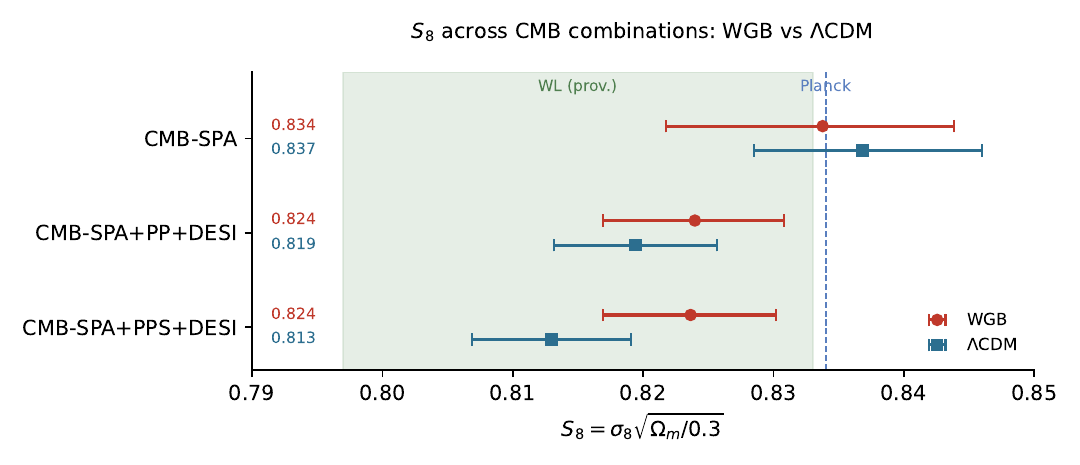}
 \caption{{\it
Marginalized $68\%$ and $95\%$ credible intervals for
$S_8\equiv\sigma_8\sqrt{\Omega_m/0.3}$ inferred from the three dataset
combinations including CMB observations, for the WGB and $\Lambda$CDM models.
Once geometric information (BAO and supernovae) is included (Runs~3 and~5),
WGB predicts a slightly higher value of $S_8$ than $\Lambda$CDM, reflecting the
enhanced growth associated with the preferred phantom branch. In the CMB-only
analysis (Run~2) the two models give compatible values, with a broader WGB
posterior reflecting the $\Cn$-$H_0$ degeneracy.
\label{fig:s8forest}}}
\end{figure}

\begin{figure}[ht]
  \centering
  \includegraphics[width=\columnwidth]{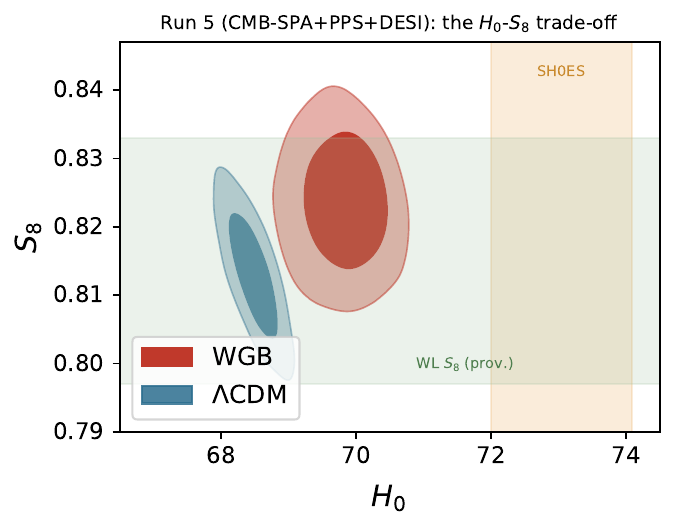}
  \caption{{\it
Marginalized $68\%$ and $95\%$ credible regions in the
$H_0$-$S_8$ plane for the full dataset combination (Run~5), comparing the
WGB (filled contours) and $\Lambda$CDM (open contours) models. Relative to
$\Lambda$CDM, WGB shifts the preferred solution toward higher values of
$H_0$, bringing it closer to the SH0ES determination, while simultaneously
predicting a modest increase in $S_8$. The figure illustrates the characteristic
$H_0$-$S_8$ trade-off of the WGB scenario.
\label{fig:s8h0}}}
\end{figure}

Applying the $Q_{\rm DMAP}$
estimator to the uncalibrated dataset combination
(CMB-SPA+PP+DESI), WGB reduces the tension from $4.74\sigma$ in
$\Lambda$CDM to $3.74\sigma$, corresponding to a relief of
$1.00\sigma$. The Gaussian estimator leads to the same conclusion, giving
$4.57\sigma\rightarrow3.56\sigma$, or a relief of $1.01\sigma$. Their close
agreement confirms that the $H_0$ posteriors are well approximated by
Gaussian distributions. Both estimators therefore indicate a consistent
alleviation of approximately $1\sigma$. At the MAP of the SH0ES-augmented
fit, the WGB solution settles at $H_0=69.70\kmsmpc$ and $\Cn=0.38$, very close
to the Run~5 best fit, with the SH0ES likelihood alone contributing
$\chi^2_{\rm SH0ES}=10.3$. This improvement is nevertheless only partial. WGB
leaves a residual ${\sim}3.7\sigma$ discrepancy with the local determination, as expected for a
purely late-time mechanism that raises $H_0$ without modifying the sound
horizon.

The corresponding behaviour of the clustering amplitude is shown in
Fig.~\ref{fig:s8forest}. Here the picture is more nuanced. For the full dataset
combination, WGB predicts a slightly larger value of $S_8$ than
$\Lambda$CDM ($0.824$ versus $0.813$ in Run~5, and $0.824$ versus $0.819$ in
Run~3), representing the modest growth-tension cost associated with the higher
value of $H_0$. In the CMB-only analysis the two models give compatible values
($S_8=0.834^{+0.010}_{-0.012}$ versus $0.837\pm0.009$). Compared with the
recent weak-lensing and clustering compilation,
$S_8=0.819\pm0.007$~\cite{Wright:2025xka,Stolzner:2025htz}, the Run~5 WGB
value lies within $0.5\sigma$, so that the price paid in $S_8$ for the
partial relief of the $H_0$ tension remains small in absolute terms.

\begin{figure}[ht]
  \centering
  \includegraphics[width=\columnwidth]{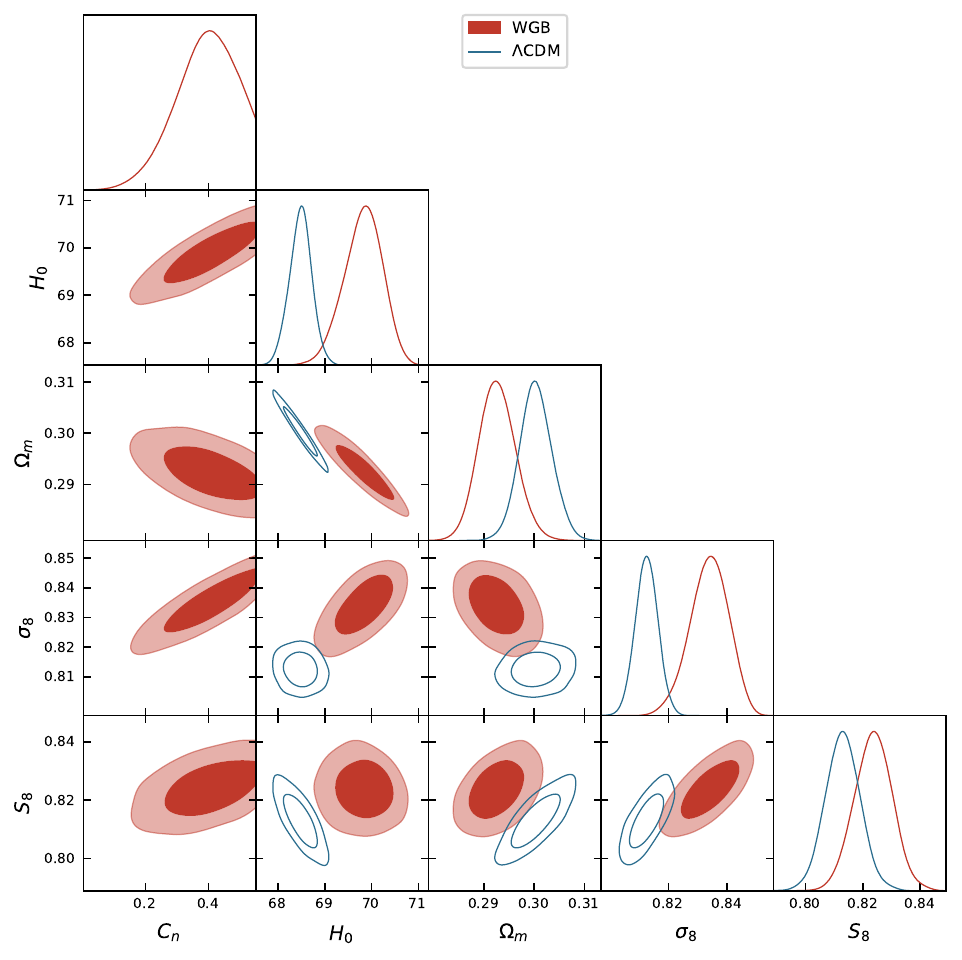}
  \caption{{\it
Marginalized $68\%$ and $95\%$ posterior distributions for the full dataset
combination (Run~5), comparing the WGB (filled contours) and
$\Lambda$CDM (open contours) models. The two-dimensional posteriors reveal
the strong positive correlations of $\Cn$ with both $H_0$ and $S_8$,
demonstrating how a positive WGB amplitude simultaneously increases the
late-time expansion rate and the growth of cosmic structure.
\label{fig:tri5}}}
\end{figure}

\begin{figure*}[t]
  \centering
  \includegraphics[width=\textwidth]{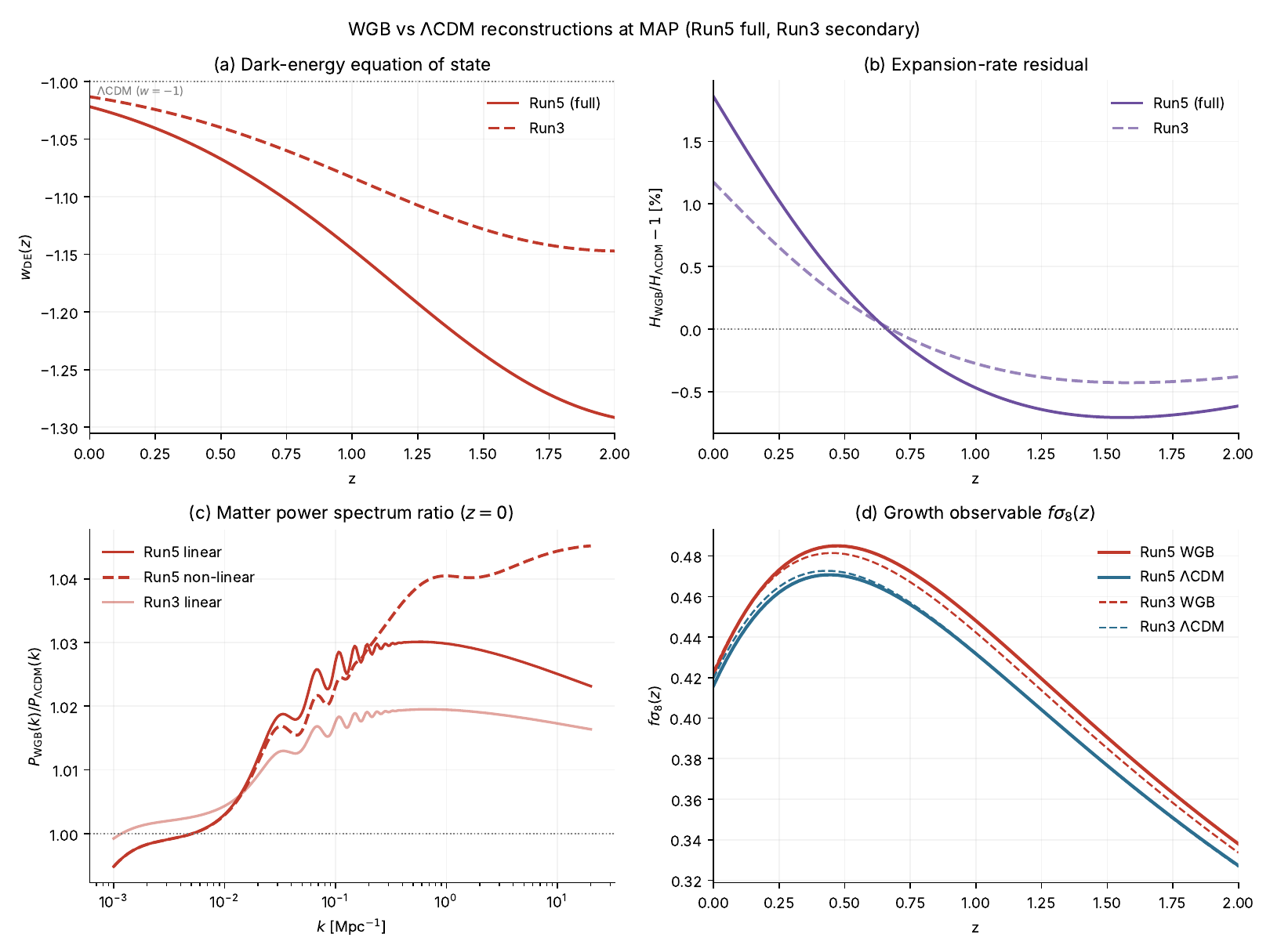}
 \caption{{\it
Reconstructed background and growth observables for the WGB model, evaluated
at the MAP points of Runs~5 (solid curves) and~3 (dashed curves), and compared
with the corresponding $\Lambda$CDM predictions. Panel~(a) shows the effective
dark-energy equation of state, $w_{\rm DE}(z)$, which evolves smoothly into the
phantom regime at late times and approaches $-1$ toward higher redshift.
Panel~(b) presents the fractional expansion-rate difference,
$H_{\rm WGB}/H_{\Lambda{\rm CDM}}-1$, displaying the enhanced expansion rate
today, the crossover near $z\simeq0.66$, and the slight suppression at
intermediate redshifts that enables a higher inferred value of $H_0$ while
preserving the acoustic scale. Panel~(c) shows the ratio of the WGB to
$\Lambda$CDM matter power spectra at $z=0$, for both the linear and nonlinear
predictions, while panel~(d) displays the corresponding growth observable
$f\sigma_8(z)$. The preferred WGB solution therefore combines a late-time
increase in the expansion rate with a modest enhancement of matter clustering,
illustrating the characteristic $H_0$-$S_8$ trade-off of the model.
\label{fig:background}}}
\end{figure*}

The joint behaviour of the two tensions is presented in
Fig.~\ref{fig:s8h0}, which highlights the trade-off between the increase in
$H_0$ and the modest enhancement of $S_8$ for the preferred WGB solution. The
full multidimensional posterior distribution is shown in
Fig.~\ref{fig:tri5}, illustrating the parameter degeneracies responsible for
this behaviour.

\section{Reconstructed observables}\label{sec:recon}

The statistical analysis of the previous section established that the WGB model
is favoured by the full cosmological dataset. We now investigate the
physical origin of this preference by reconstructing the principal background
and perturbation observables at the MAP points of the two fully constrained
dataset combinations (Runs~3 and~5), using the modified \texttt{CLASS}
implementation described in subsection \ref{sec:class}. For the CMB and matter-power
spectra we additionally propagate the $68\%$ posterior uncertainty by
re-evaluating the Boltzmann solver over $200$ samples drawn from each MCMC
chain.

\begin{figure*}[t]
  \centering
  \subfloat[CMB temperature]{\includegraphics[width=0.32\textwidth]{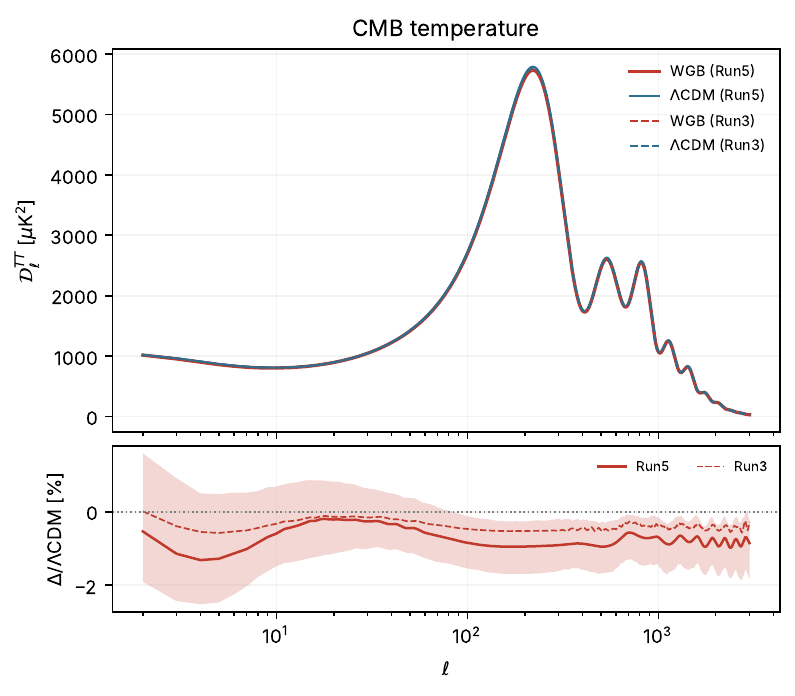}}\hfill
  \subfloat[CMB lensing]{\includegraphics[width=0.32\textwidth]{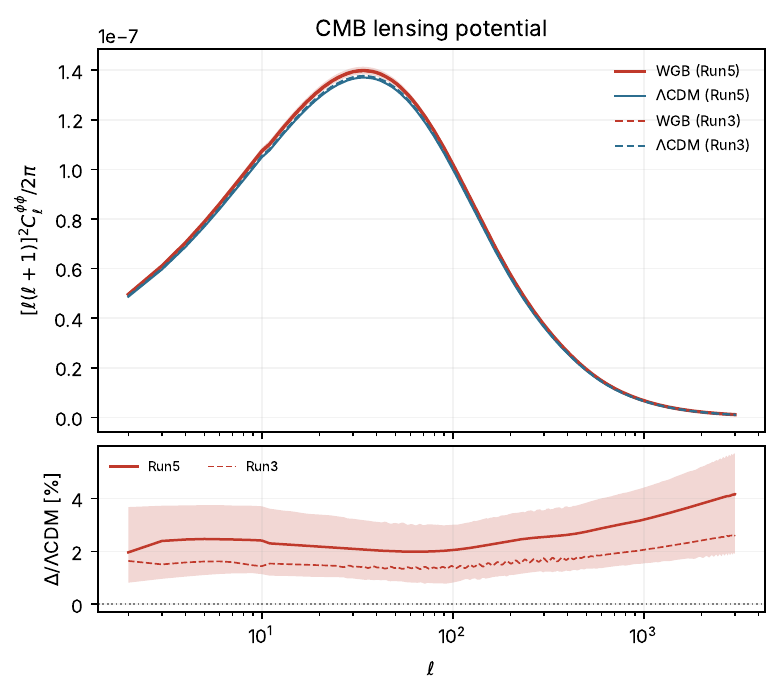}}\hfill
  \subfloat[Matter power]{\includegraphics[width=0.32\textwidth]{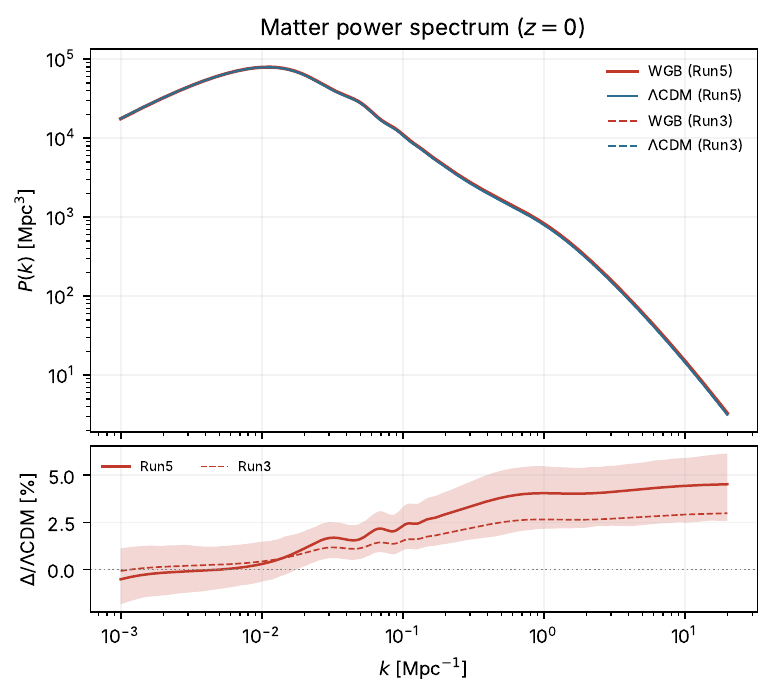}}
\caption{{\it
Reconstructed CMB and matter-power spectra for the WGB model, evaluated at
the MAP points of Runs~5 (solid curves, with the $68\%$ posterior band) and~3
(dashed curves), and compared with the corresponding $\Lambda$CDM
predictions. Panel (a) shows the lensed CMB temperature spectrum
$\mathcal{D}_\ell^{TT}$, panel (b) the CMB lensing spectrum
$\mathcal{D}_\ell^{\phi\phi}$, and panel (c) the matter power spectrum
$P(k)$. The lower panels display the percentage difference relative to
$\Lambda$CDM. While the primary CMB temperature spectrum is affected only at
the sub-percent level, WGB predicts enhanced CMB lensing and increased
small-scale matter clustering, consistent with the higher inferred value of
$S_8$.
\label{fig:spectra}}}
\end{figure*}

\subsection{Expansion history}

The reconstructed background evolution is shown in
Fig.~\ref{fig:background}. The WGB equation of state (panel a) remains smoothly
phantom, evolving from $w_{\rm DE}(0)\simeq-1.02$ today to a minimum of
$\simeq-1.30$ at $z\simeq2.2$ in Run~5, near the peak of cosmic star
formation. The evolution is milder in Run~3,
reflecting its smaller value of $\Cn$, and gradually approaches the
$\Lambda$CDM behaviour at high redshift as the star-formation source
$\psi(z)$ becomes negligible. This phantom evolution translates into an
approximately $1.9\%$ higher expansion rate today (panel b), crossing below the
$\Lambda$CDM prediction near $z\simeq0.66$ and reaching a
${\sim}0.7\%$ deficit at $z\simeq1.5$. This characteristic behaviour reflects
the exchange between enhanced late-time acceleration and the fixed
early-Universe acoustic scale.

The origin of the higher $H_0$ is transparent in terms of the acoustic scale.
Because the WGB modification is sourced by the cosmic star-formation history,
it is negligible before $z\sim10$ and therefore leaves the sound horizon at
decoupling, $r_s(z_\star)$, unchanged with respect to $\Lambda$CDM. The phantom
equation of state ($w_{\rm DE}<-1$) suggests a lower value for the dark-energy density in the
past, suppressing $H(z)$ at intermediate redshifts. The
${\sim}0.7\%$ deficit visible in Fig.~\ref{fig:background}(b) at
$z\gtrsim1$ would increase the comoving distance to last scattering,
$D_A(z_\star)$. Since the CMB constrains the angular acoustic scale,
$\theta_\star=r_s(z_\star)/D_A(z_\star)$, to sub-percent precision, the fit
compensates by increasing the late-time expansion rate, namely $H_0$, which
restores $D_A(z_\star)$. The
change of sign in the expansion-rate residual around $z\simeq0.66$ is the
characteristic signature of this compensation: a deficit in $H(z)$ when dark
energy was less important, balanced by an excess today while preserving
$\theta_\star$. This is the same geometric mechanism encountered in other
effective phantom scenarios
\cite{ElZant:2019,DiValentino:2016}, although here it arises naturally from the
WGB framework without introducing any new early-Universe physics.

\subsection{Growth of cosmic structure}

The reconstructed CMB and matter-power spectra are presented in
Fig.~\ref{fig:spectra}. The primary CMB temperature anisotropies are almost
indistinguishable from those of $\Lambda$CDM (panel a). The differences remain
at the sub-percent level and are confined to a mild low-$\ell$ deficit,
reaching approximately $-1.3\%$ near $\ell\simeq4$ because of the suppressed
late-time integrated Sachs-Wolfe effect, together with very small peak shifts,
all well within the posterior uncertainty.

\begin{figure}[ht]
  \centering
  \includegraphics[width=\columnwidth]{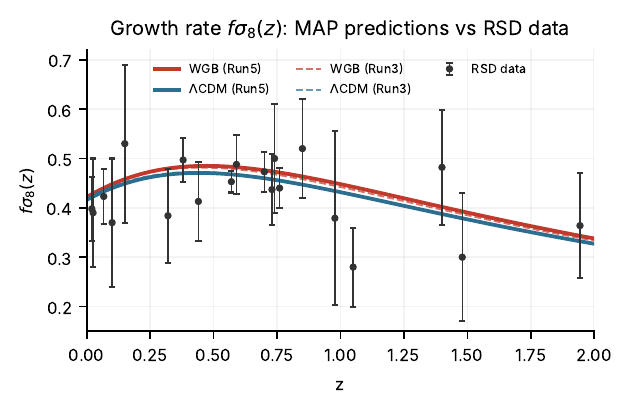}
 \caption{{\it
Predicted growth rate $f\sigma_8(z)$ for the WGB and $\Lambda$CDM models,
evaluated at the MAP points of Runs~5 (solid curves) and~3 (dashed curves),
and compared with the 20-point redshift-space distortion compilation of
Ref.~\cite{Avila2022}. The RSD measurements are not included in the likelihood
and therefore provide an independent validation of the reconstructed growth
history. WGB predicts a modest enhancement of structure growth relative to
$\Lambda$CDM, while remaining fully consistent with the current observational
scatter.
\label{fig:fs8}}}
\end{figure}

The impact of WGB is instead visible in the late-time clustering observables.
The CMB lensing spectrum, shown in panel~(b), is enhanced by approximately
$2$-$4\%$, while the matter power spectrum in panel~(c) increases by up to
${\sim}4.5\%$ on nonlinear scales
($k\gtrsim1\,{\rm Mpc}^{-1}$), with the posterior band excluding zero. This
combination of enhanced lensing and enhanced clustering is the direct
consequence of the phantom background evolution and explains both the slightly
larger value of $S_8$. The enhancement is nevertheless compatible with the
ACT~DR6 and SPT-3G lensing reconstructions, whose $\chi^2$ is essentially
unchanged with respect to $\Lambda$CDM (subsection \ref{sec:modelcomp}): the
growth cost of the higher $H_0$ is absorbed within the current lensing
uncertainties.

As an independent validation, we compare the predicted growth rate
$f\sigma_8(z)$ with the compilation of $20$ redshift-space distortion
measurements of Ref.~\cite{Avila2022}, the same dataset used in the companion
late-time analysis~\cite{Petronikolou:2025mlm}. These measurements are
\emph{not} included in the likelihood and therefore constitute a genuine
prediction rather than a fit. Following
Ref.~\cite{Petronikolou:2025mlm}, the measurements are treated as independent
and no Alcock-Paczynski correction is applied. The reconstructed growth
history is shown in Fig.~\ref{fig:fs8}. WGB predicts a growth rate lying
slightly above the $\Lambda$CDM expectation, although both remain comfortably
within the observational scatter. The corresponding point-by-point comparison
is reported in Appendix \ref{app:pointbypoint}. The RSD compilation exhibits a mild preference for
$\Lambda$CDM, with
$\chi^2=11.4\,(8.6)$ for WGB ($\Lambda$CDM) in Run~5, corresponding to
$\Delta\chi^2(\mathrm{WGB}-\Lambda\mathrm{CDM})=+2.8$
($+1.6$ for Run~3), while both models satisfy
$\chi^2/N<0.6$. Redshift-space distortions therefore provide an independent
growth probe that mildly disfavours the WGB growth enhancement, illustrating
the price paid for the partial alleviation of the $H_0$ tension.

\section{Discussion and conclusions}\label{sec:conclusions}

One of the central open questions in modern cosmology is whether the
observed late-time tensions point to new physics beyond $\Lambda$CDM paradigm, or merely
reflect residual systematic effects. Among the many proposed extensions,
late-time dark-energy scenarios are particularly attractive because they can
modify the recent expansion history while preserving the remarkable success of
the standard cosmological model at early times. Wald-Gauss-Bonnet (WGB)
topological dark energy provides a novel realization of this idea. Rather than
introducing new fundamental fields or modifying the gravitational field
equations, the model attributes the effective dark-energy sector to the
thermodynamic properties of the cosmic apparent horizon, with the additional
contribution determined by the cosmic history of black-hole formation and
mergers.

In this work we have performed the first complete early- and late-Universe
confrontation of WGB cosmology. Building on the late-time analysis of
Ref.~\cite{Petronikolou:2025mlm}, we implemented the Model~II realization in a
modified version of the \texttt{CLASS} Boltzmann solver, consistently evolved
the background and perturbations through recombination, and constrained the
model against complementary combinations of \textit{Planck}, ACT~DR6,
SPT-3G, DESI~DR2 BAO, and Pantheon+ supernova observations. This constitutes
the first CMB test of a cosmological scenario in which dark energy originates
from the interplay between horizon thermodynamics and astrophysical
black-hole evolution.

Our analysis shows that current cosmological observations are fully compatible
with the WGB framework and, more importantly, reveal a preference for a
non-trivial WGB contribution once early- and late-Universe probes are analysed
jointly. For the full dataset combination we obtain
$\Cn=0.387^{+0.118}_{-0.072}$, corresponding to a more than
$3\sigma$ preference for the phantom branch of the model. Relative to
$\Lambda$CDM, the fit improves by
$\Delta\chi^2=-10.3$, with
$\Delta\mathrm{AIC}=-8.3$, $\Delta\mathrm{BIC}=-2.6$ and
$\Delta\mathrm{DevIC}=-9.3$, while the Bayes factor,
$\ln B_{10}=3.2$, indicates strong evidence for a non-zero WGB amplitude. Significantly, this preference emerges only after
including the CMB information. Late-time observations alone remain statistically
compatible with both WGB and $\Lambda$CDM, highlighting the crucial role of
precision early-Universe measurements in testing physically motivated
dark-energy models.

The preferred WGB solution raises the Hubble constant from
$68.5$ to $69.8\,\mathrm{km\,s^{-1}\,Mpc^{-1}}$, reducing the Hubble tension by
approximately $1\sigma$ (from $4.7\sigma$ to $3.7\sigma$) without introducing any modification to early-Universe
physics. The underlying mechanism is entirely geometric: because the WGB
contribution becomes relevant only after the onset of significant cosmic
star formation, the sound horizon remains unchanged while the late-time
expansion history is modified through a smoothly phantom equation of state.
The reconstructed observables demonstrate that this mechanism leaves the
primary CMB temperature anisotropies essentially unchanged while enhancing CMB
lensing and small-scale matter clustering. Consequently, the partial relief of
the $H_0$ tension is accompanied by a mild increase in $S_8$, which remains
within $0.5\sigma$ of current weak-lensing and clustering determinations and
is mildly disfavoured by the independent RSD measurements.

These results establish WGB cosmology as a viable and testable alternative to
conventional late-time dark-energy models. Equally importantly, they illustrate
that topological and thermodynamic properties of gravity can lead to observable
cosmological consequences without introducing additional propagating degrees of
freedom or altering the Einstein field equations. Although the present evidence
depends primarily on the combined constraining power of
current CMB and distance-ladder data, the framework has now reached the stage
where it can be confronted quantitatively with precision cosmological
observations on the same footing as more conventional dark-energy scenarios.

Future data will provide decisive tests of this picture. Improved CMB lensing,
Stage-IV weak-lensing and galaxy surveys, together with increasingly precise
measurements of black-hole populations and merger rates, will tighten the
constraints on the effective parameter $\Cn$ and may eventually disentangle its
astrophysical origin through the factor $C$. Whether the present preference for
a non-zero WGB contribution ultimately strengthens or disappears, this work
demonstrates that horizon thermodynamics and spacetime topology provide a
predictive cosmological framework whose observational consequences are now
accessible to precision cosmology.

\begin{acknowledgments}
This work was carried out within the CosmoVerse Compilation Group (CCG)
initiative, and its results will be presented in the forthcoming CosmoVerse white
paper~\cite{CosmoVerse2025}. The authors would like to acknowledge the
contribution of the LISA CosWG, and of COST Actions CA21136 ``Addressing
observational tensions in cosmology with systematics and fundamental physics
(CosmoVerse)'', CA21106 ``COSMIC WISPers in the Dark Universe: Theory,
astrophysics and experiments (CosmicWISPers)'', and CA23130 ``Bridging high and
low energies in search of quantum gravity (BridgeQG)''. Numerical computations
were performed on the computing facilities of the National Observatory of Athens. AP acknowledges the support from FONDECYT Grant 1240514.
\end{acknowledgments}

\appendix

\section{Point-by-point likelihood comparison}
\label{app:pointbypoint} 

To identify which observations drive the differences between the WGB and
$\Lambda$CDM fits, we compare the individual likelihood contributions for
each data point. Rather than considering only the total $\chi^2$, this
point-by-point decomposition reveals which measurements favour one model
over the other and therefore provides additional insight into the origin of
the overall fit improvement discussed in subsection \ref{sec:modelcomp}.

Table~\ref{tab:rsd} lists the individual $\chi^2$ contributions
for the principal datasets considered in this work, together with the
corresponding differences between WGB and $\Lambda$CDM. Positive values of
$\Delta\chi^2$ indicate a preference for $\Lambda$CDM, whereas negative
values favour the WGB model.

\begin{table}[ht]
\caption{Compilation of the redshift-space distortion measurements
($f\sigma_8$) from Ref.~\cite{Avila2022}, together with the corresponding
MAP predictions of the WGB and $\Lambda$CDM models for the full dataset
combination (Run~5). These measurements are not included in the likelihood and
therefore provide an independent validation of the reconstructed growth
history. Following Ref.~\cite{Petronikolou:2025mlm}, the data points are
treated as independent and no Alcock-Paczynski correction is applied.
\label{tab:rsd}}
\begin{ruledtabular}\begin{tabular}{cccc}
$z$ & $f\sigma_8^{\rm obs}$ & $f\sigma_8^{\rm WGB}$ & $f\sigma_8^{\Lambda{\rm CDM}}$ \\ \hline
0.020 & $0.398\pm0.065$ & 0.428 & 0.421 \\
0.025 & $0.390\pm0.110$ & 0.429 & 0.423 \\
0.067 & $0.423\pm0.055$ & 0.440 & 0.433 \\
0.100 & $0.370\pm0.130$ & 0.448 & 0.440 \\
0.150 & $0.530\pm0.160$ & 0.458 & 0.449 \\
0.320 & $0.384\pm0.095$ & 0.480 & 0.467 \\
0.380 & $0.497\pm0.045$ & 0.483 & 0.470 \\
0.440 & $0.413\pm0.080$ & 0.485 & 0.471 \\
0.570 & $0.453\pm0.022$ & 0.483 & 0.468 \\
0.590 & $0.488\pm0.060$ & 0.482 & 0.467 \\
0.700 & $0.473\pm0.041$ & 0.476 & 0.460 \\
0.730 & $0.437\pm0.072$ & 0.474 & 0.458 \\
0.740 & $0.500\pm0.110$ & 0.473 & 0.457 \\
0.760 & $0.440\pm0.040$ & 0.472 & 0.455 \\
0.850 & $0.520\pm0.100$ & 0.463 & 0.447 \\
0.978 & $0.379\pm0.176$ & 0.450 & 0.434 \\
1.050 & $0.280\pm0.080$ & 0.442 & 0.426 \\
1.400 & $0.482\pm0.116$ & 0.402 & 0.387 \\
1.480 & $0.300\pm0.130$ & 0.393 & 0.379 \\
1.944 & $0.364\pm0.106$ & 0.343 & 0.332 \\
\hline
\multicolumn{2}{l}{$\chi^2$ (Run~5, $N=20$)} & 11.38 & 8.60 \\
\multicolumn{2}{l}{$\chi^2$ (Run~3, $N=20$)} & 10.39 & 8.81 \\
\end{tabular}\end{ruledtabular}\end{table}

\section{Numerical implementation details}\label{app:numerics}

The WGB background evolution described in
subsection \ref{sec:class} is implemented in the
\texttt{class\_wgb} fork of \texttt{CLASS}. In this appendix we provide
additional details concerning the numerical evaluation of the
star-formation contribution and the rejection of unphysical parameter
configurations.

The equation of state in Eq.~(\ref{eq:wdea}) contains the integral
\begin{equation}
I(a)=\int_a^1 \frac{\hat\psi(a')}{a'}\,da',
\end{equation}
which must be evaluated repeatedly during the background integration. A
direct implementation that recomputed this quantity at every integration
step would duplicate a substantial amount of numerical work. In an initial
version of the code, it also led to a memory leak because the integration
workspace was repeatedly allocated without being released. We therefore
evaluate $I(a)$ once on the background sampling grid and construct a spline
interpolation that is queried throughout the integration. This allows
$w_{\rm DE}(a)$ and $\rho_{\rm DE}(a)$ to be evolved within a single
background pass, substantially improving both numerical efficiency and
stability.

A second numerical issue arises for sufficiently large positive values of
$\Cn$. In this regime, the denominator of Eq.~(\ref{eq:wdea}), namely
$D(a)=3(1-\Omega_{m0})-24\Cn\int_a^1\hat\psi(a')/a'\,da'$,
may vanish and subsequently change sign within the integration range
[cf.~Eq.~(\ref{eq:cncrit})]. Such
configurations lead to a negative effective dark-energy density and can
render the comoving angular-diameter distance ill defined,
$D_A\leq0$. We therefore impose an explicit \texttt{class\_test} that
monitors this denominator during the background evolution. Parameter
combinations that violate the positivity condition are identified as
unphysical and rejected before the perturbation and spectral calculations
are performed. Together with the prior $\Cn\in[-10,0.55]$ adopted in
subsection \ref{sec:class}, this procedure ensures that every accepted sample
corresponds to a well-defined and physically admissible expansion history.

The best-fit $\chi^2$ values decomposed by likelihood for
the two fully constrained dataset combinations are presented in
Table~\ref{tab:chi2} above. The individual CMB blocks are reported as returned
by the corresponding likelihoods and sum to the total $\chi^2$ used in all
information criteria (subsection \ref{sec:mcmc}).

\section{Per-dataset posterior behaviour}\label{app:triangles}

The parameter constraints presented in the main text correspond to five
different dataset combinations (Runs~1-5). Since the scientific conclusions
are driven by the fully constrained combination (Run~5), whose marginalized
posterior distributions are shown in Fig.~\ref{fig:tri5}, we do not reproduce
the corresponding corner plots for the remaining runs.

Their posterior behaviour is nevertheless instructive. The two late-time-only
analyses (Runs~1 and~4) exhibit broad posteriors centred on mildly negative
values of $\Cn$ and consistent with zero at $95\%$ credibility, reflecting the
limited constraining power of late-time distance measurements alone. The
intermediate combination (Run~3) already favours positive values of $\Cn$,
although only at the ${\sim}2\sigma$ level, and displays parameter degeneracies qualitatively
similar to those of the full Run~5 analysis. The CMB-only combination (Run~2)
is qualitatively different. There, the nearly perfect
$\Cn$-$H_0$ degeneracy (sample correlation
$\rho\simeq0.97$) drives the posterior toward the upper prior boundary,
demonstrating that the apparent preference for positive $\Cn$ is entirely a
degeneracy effect rather than evidence for a non-zero WGB contribution.

In summary, the evolution of the posterior distributions across the five dataset
combinations is fully consistent with the parameter constraints discussed in
subsection \ref{sec:constraints}. The inclusion of complementary geometric probes
progressively breaks the $\Cn$-$H_0$ degeneracy, leading to the statistically
significant preference for a positive WGB amplitude only in the complete
early- and late-Universe dataset combination.

\bibliography{refs}

\end{document}